\documentclass[useAMS,usenatbib]{mn2e}
\usepackage{epsfig}

\newcommand{\ltsima} {$\; \buildrel < \over \sim \;$}
\newcommand{\gtsima} {$\; \buildrel > \over \sim \;$}
\newcommand{\lta} {\lower.5ex\hbox{\ltsima}}
\newcommand{\gta} {\lower.5ex\hbox{\gtsima}}

\title[GRB110920A: extremely narrow spectrum]{Extremely narrow spectrum of GRB110920A: further evidence for localised, subphotospheric dissipation}
\author[S. Iyyani et al.]{S. Iyyani$^{1,2,3,4}$\thanks{email: shabuiyyani@particle.kth.se},  
F. Ryde$^{1,2}$,  B. Ahlgren$^{1,2}$, J. M. Burgess$^{1,2}$, J. Larsson$^{1,2}$, A. Pe'er$^{7}$, \newauthor C. Lundman$^{5}$, M. Axelsson$^{2,3,8}$, S. McGlynn$^{6}$\\
$^{1}$Department of Physics, KTH Royal Institute of Technology, AlbaNova University Center, SE-106 91 Stockholm, Sweden\\ 
$^{2}$The Oskar Klein Centre for Cosmoparticle Physics, Stockholm, Sweden\\ 
$^{3}$Department of Physics, Stockholm University, Sweden\\ 
$^{4}$Erasmus Mundus Joint Doctorate in Relativistic Astrophysics\\
$^{5}$Department of Physics, Columbia University, New York, USA\\
$^{6}$University College Dublin, Ireland\\
$^{7}$University College Cork, Ireland\\
$^{8}$Department of Astronomy, Stockholm University, Sweden\\
}

\begin{document}

\date{Accepted... Received...; in original form ...}

\pagerange{\pageref{firstpage}--\pageref{lastpage}} \pubyear{2015}

\maketitle

\label{firstpage}

\begin{abstract}
Much evidence points towards that the photosphere in the relativistic outflow in GRBs plays an important role in shaping the observed MeV spectrum. However, it is unclear whether the spectrum is fully produced by the photosphere
or whether a substantial part of the spectrum is added by processes far above the photosphere. Here we make a detailed study of the $\gamma-$ray emission from single pulse GRB110920A which has a spectrum that becomes extremely narrow towards the end of the burst. 
We show that the emission can be interpreted as Comptonisation of thermal photons by cold electrons in an unmagnetised outflow at an optical depth of $\tau \sim 20$. The electrons receive their energy by a local dissipation occurring close to the saturation radius.   
The main spectral component  of GRB110920A and its evolution is thus, in this interpretation, fully explained by the emission from the photosphere including localised dissipation at high optical depths. 
\end{abstract}


\section{Introduction}

Photospheric emission has currently become the pivot of the study of radiation mechanisms in gamma-ray bursts (GRBs). In the relativistic fireball model (e.g. \citet{Meszaros2006}),  large amounts of energy is injected at the base of the flow where the  optical depth is huge and  the photon field  gets efficiently thermalised. As the fireball expands the optical depth of the plasma eventually reaches unity and the  photons are able to decouple from the plasma. The region from where the photons escape is called the photosphere. 

Two main perspectives of photospheric emission models  are currently discussed:
In the first perspective, the observed spectrum is the result of emission from two different emission zones, the photosphere and an optically-thin region \citep{Meszaros&Rees2000, Zhang&Meszaros2002, Daigne&Mochkovitch2002}. 
The emission from the photosphere is assumed to be close to a blackbody, implying a passive flow below the photosphere. The dissipation of the kinetic energy (or alternatively Poynting flux) that takes place in the optically-thin region results in non-thermal processes such as synchrotron emission \citep{Tavani1996}  and/or inverse Compton scattering \citep{Ghisellini2000} producing the non-thermal part of the spectrum \citep{Meszaros2002,Deng2014}.  
Based on such an  interpretation, observed MeV spectra have been fitted with a blackbody and a power law (over the {\it CGRO} BATSE energy range; \cite{Ryde2004, Ryde2005}, \cite{Ryde&Pe'er2009}) or a blackbody combined with a Band function (over the {\it Fermi} energy range \cite{Guiriec2011, Axelsson2012, Iyyani2013,Guiriec2013}). 

In the second perspective, the entire spectrum is due to emission in the vicinity of the photosphere; additional processes cause the spectrum from the photosphere to differ from a blackbody \citep{Rees&Meszaros2005}. Energy dissipation  below the photosphere (subphotospheric dissipation)  can cause significant broadening of the thermal component in the flow. The dissipation can be imagined to be local, at a certain position in the flow \citep{Rees&Meszaros2005, Pe'er2005, Beloborodov2010},
or be continuous, having an effect through out the flow \citep{Giannios2012, Beloborodov2013}. Moreover, the photosphere  is, in general, not expected to be a surface, defined by a single photospheric radius, but rather an extended volume. Variations in the temperature due to expansion \citep{Beloborodov2011} as well as high latitude effects on the Lorentz boost of the emission \citep{Abramowicz1991} will cause the observer to see a multi-temperature emission, significantly broadening the spectrum (Pe'er 2008, see also Goodman1986, Lundman et al. 2013). 

First observational evidence for photospheric, spectral broadening was given by GRB 090902B, during which there was an onset of a broadening mechanism: Initially the spectrum is well described by a (narrow) multicolour blackbody, not much broader than a blackbody \citep{Ryde2010, Zhang2011}. Later, the burst spectrum evolves and gets broader, now resembling a more typical (broad) spectral shape \citep{Ryde2011}. 
In this paper, we study the $160$ s long pulse in GRB110920A, which is very well suited for time resolved spectral analysis. Initial analyses \citep{McGlynn2012, Shenoy2013} have shown that the spectra  are hard and part of the spectrum has been suggested to be of  photospheric origin. Therefore, this burst is a strong candidate to detect the photosphere and to study its properties. We find that the spectrum and its evolution gives additional, strong support for the existence of energy dissipation below the photosphere, shaping the observed spectrum.  

This paper is organised as follows: \S \ref{obs} describes the observations and results of the spectral analysis of the burst; \S \ref{Phys_scenario} discusses the physical scenario resulting in the observed spectra; \S \ref{Discussion} is a general discussion about the various properties of the burst and finally \S \ref{Conclusion} gives the conclusion.

\section{Observations and spectral analysis}
\label{obs}
\begin{figure}
\begin{center}
\resizebox{84mm}{!}{\includegraphics{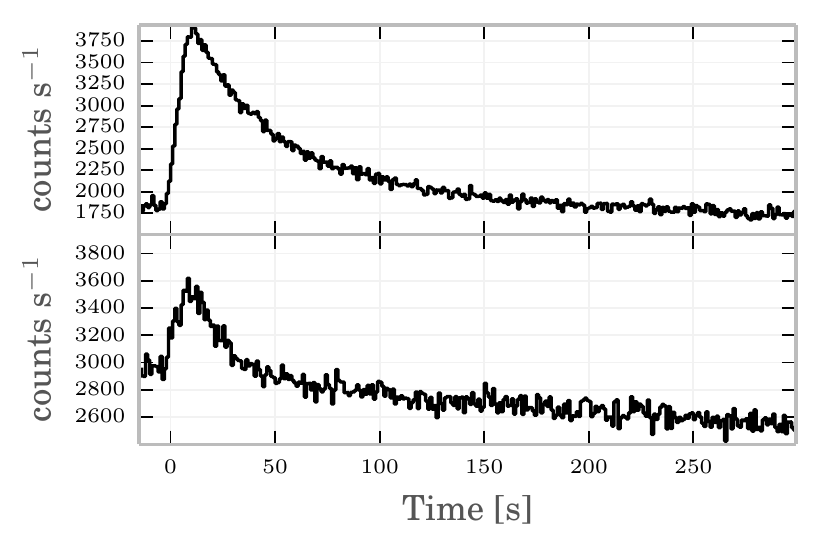}}
\caption{\small The light curve of GRB110920A as was observed by the GBM detectors: the brightest sodium iodide detector, NaI 3 (8 - 800 keV; upper panel) and BGO 0 (200 keV - 40 MeV; lower panel) are shown. }
\label{fig:lc}
\end{center}
\end{figure}

GRB110920A was observed on 20 September 2011 by the Gamma Ray Burst Monitor onboard {\it Fermi Gamma-ray Space Telescope} . It has a fluence of $1.74 \pm 1.24  \times 10^{-4} \, \rm erg\, cm^{-2}$ in the energy range $10$ keV - $40$ MeV. There is no detection of Large Area Telescope (LAT, 100 MeV - 300 GeV) and LAT Low Energy (LLE, 30 - 130 MeV) emission for this burst. No afterglow has been detected for the burst and therefore, the redshift, $z$, is unknown. The light curve of the burst is a single pulse with a $T_{90} = 161$ s, the time during which $90 \%$ of the emission is received. The duration is much longer than the average long burst, with only $\sim 5\%$ of bursts having a longer duration \citep{vonKienlin2014}.
The variability timescale is much longer than the estimated dynamical time of the burst which is of the order of $t_{\rm dyn}  \sim 0.2$ ms (see \S \ref{best_fit_param}).
Such smooth pulses enable us to follow the spectral evolution more distinctly in comparison to other bursts where there are many spikes in the light curve which makes it difficult to resolve the different episodes of emission. Figure \ref{fig:lc}  shows the composite light curve of the sodium iodide (NaI) and bismuth gallium oxide (BGO) detectors of GBM, that detected the burst \citep{Meegan2009}. 
The energy flux in the energy range $10$ keV - $40$ MeV peaks at $12$ s with $5.8 \pm 0.55 \times 10^{-6}  \rm erg\, cm^{-2}\, s^{-1}$.

A standard time resolved spectral analysis of the burst with HEASARC's spectral analysis software package XSPEC \footnote{{\scriptsize http://heasarc.nasa.gov/xanadu/xspec}} \citep{Arnaud1996}, using PHA (Pulse Height Analyser) data from NaI 0,1,3 and the BGO 0 detectors in the energy range 10 keV - 40 MeV was performed. 
The burst was analysed for its duration of $0$ to $200$ s. After $60$ s there was no significant BGO emission. However, the BGO was included in the analysis from $60$ to $200$ s in order to include the upper limits of BGO emission. 
The time bins for the spectral analysis were chosen for a signal-to-noise (S/N) ratio = $60$, thereby enabling us to constrain the evolution of the spectral properties with high level of significance (see \cite{Burgess2014c}). 

In search of the spectral shape of the emission, three empirical functions for the photon flux ($N_{\rm E}(E)$, in units of $\rm photons/cm^2/s/keV$) are used to fit the spectrum:\\
(i) The Band et al. (1994) function, which consists of  two smoothly joined power-laws, and has four free parameters: 
$E_{\rm peak}$, the peak in the $\nu F_{\nu}$ spectrum,  $\alpha$ the asymptotic power law index below the peak and $\beta$ the asymptotic power-law index above the peak, apart from the normalisation.  \\
(ii) The blackbody (BB), which  has two free parameters: the  temperature, $T$, and the normalisation, $A$;
\begin{equation}
N_{\rm E} (E) = A \,\, \frac{ E^2}{\exp[E/kT]-1},
 \label{eq:BBph}
\end{equation}
\noindent
where $k$ is the Boltzmann constant.\\
(iii) The power-law (PL), which has two free parameters: the power-law index, $s$, and the normalisation, $K$:
$N(E) = K(E/E_0)^{s}$, where $E_0 = 1$ keV. 

The Band function and power law have no direct physical meaning, however the Band function can be interpreted as being the result of an emission mechanism, such as synchrotron or inverse Compton emission. Likewise, the power-law can be a valid approximation of various emission processes over a limited energy range. The blackbody can be interpreted as the thermal emission from the photosphere of the outflow. 
 
 In the following, we use different combinations of these functions to asses three general models: a single non-thermal emission component (Band function alone, \S \ref{sec:band}), a two emission-zone model (Band+BB, \S \ref{sec:bandbb}) and finally a Comptonisation model, which is characterised by two temperatures, the injected photon temparature and the electron temperature. The latter model  also includes a power law component which can be associated to an optically-thin emission, not directly associated with the main Comptonisation event (BB+BB+PL, \S \ref{bb2pow}). 

\subsection{Inability of the Band function to fit the data}

GRB spectra are typically fitted with standard empirical models consisting of a single Band function with or without an additional blackbody. For GRB110920A such functions do not produce the best fit to the data as will be shown below. 

\subsubsection{Single Band function fits}
\label{sec:band}
A Band function was fitted to the spectra of individual time bins through out the burst. This is the standard analysis that is performed in, for instance, the  {\it Fermi} GBM  \citep{GBMcatalog2014} and {\it CGRO} BATSE catalogues \citep{BATSEcatalog2013}.  The fits are usually interpreted as a single non-thermal emission component, such as synchrotron or inverse Compton emission. 

\begin{figure*}[h]
\begin{center}
\resizebox{81mm}{!}{\includegraphics{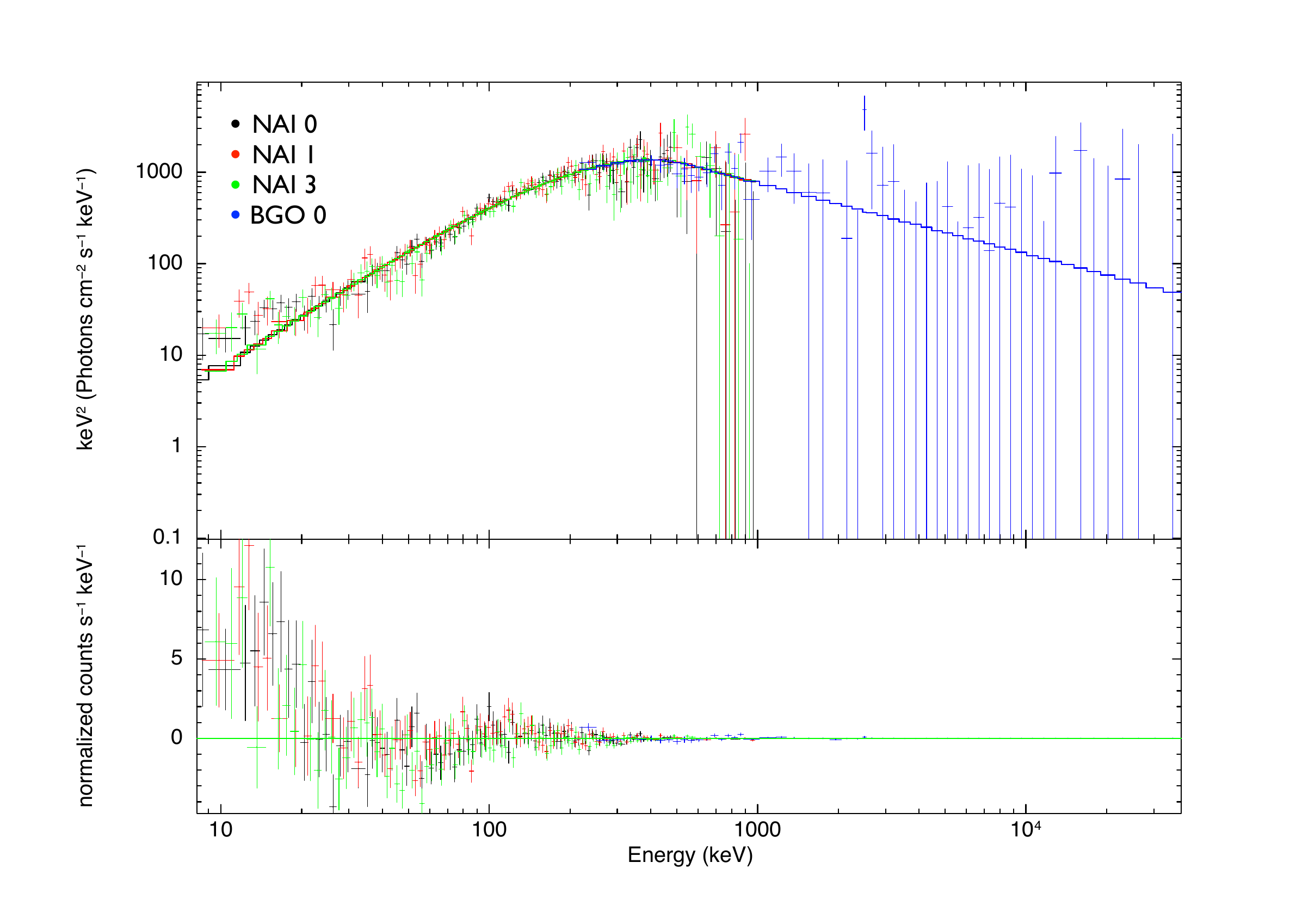}}
\resizebox{81mm}{!}{\includegraphics{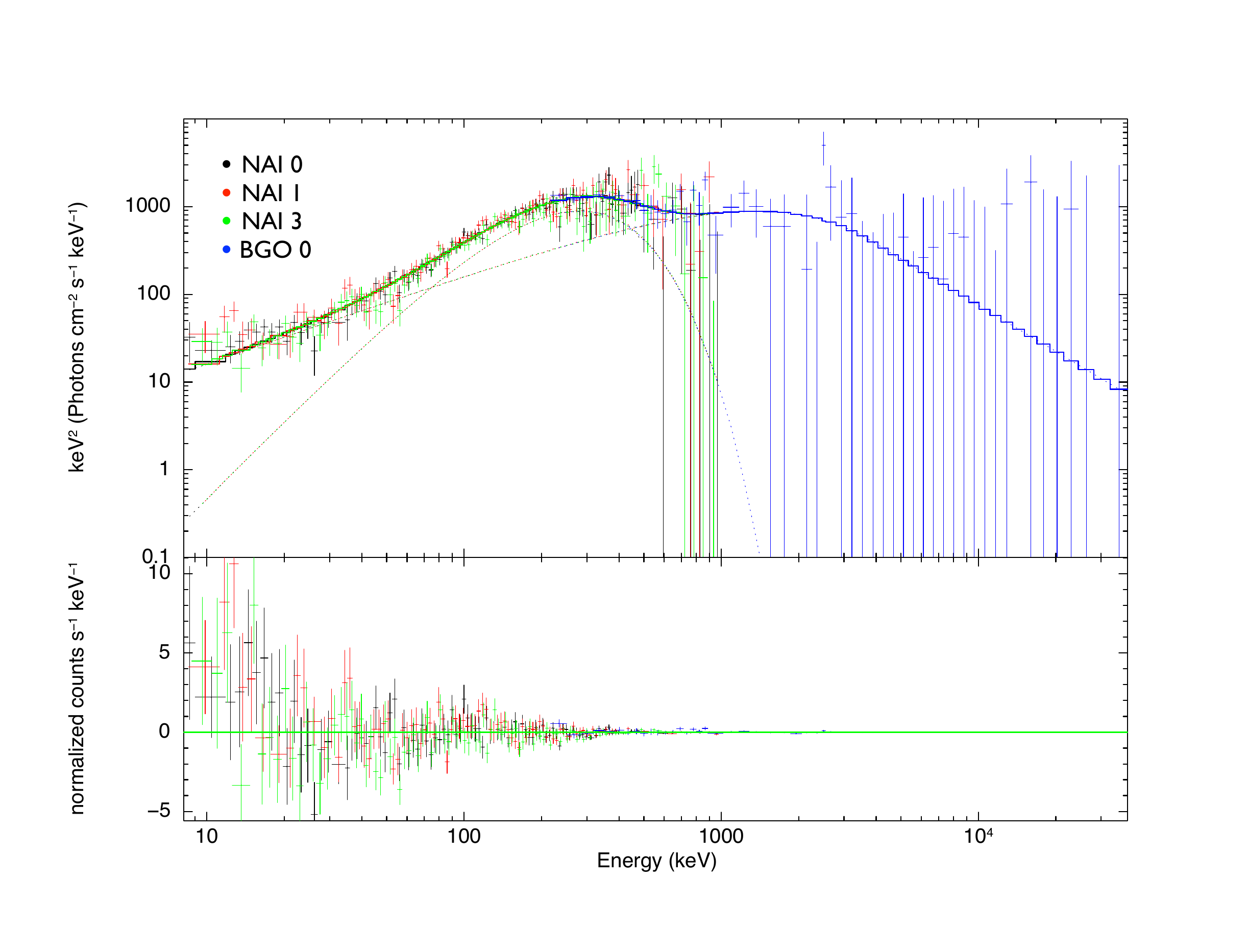}}
\resizebox{85mm}{!}{\includegraphics{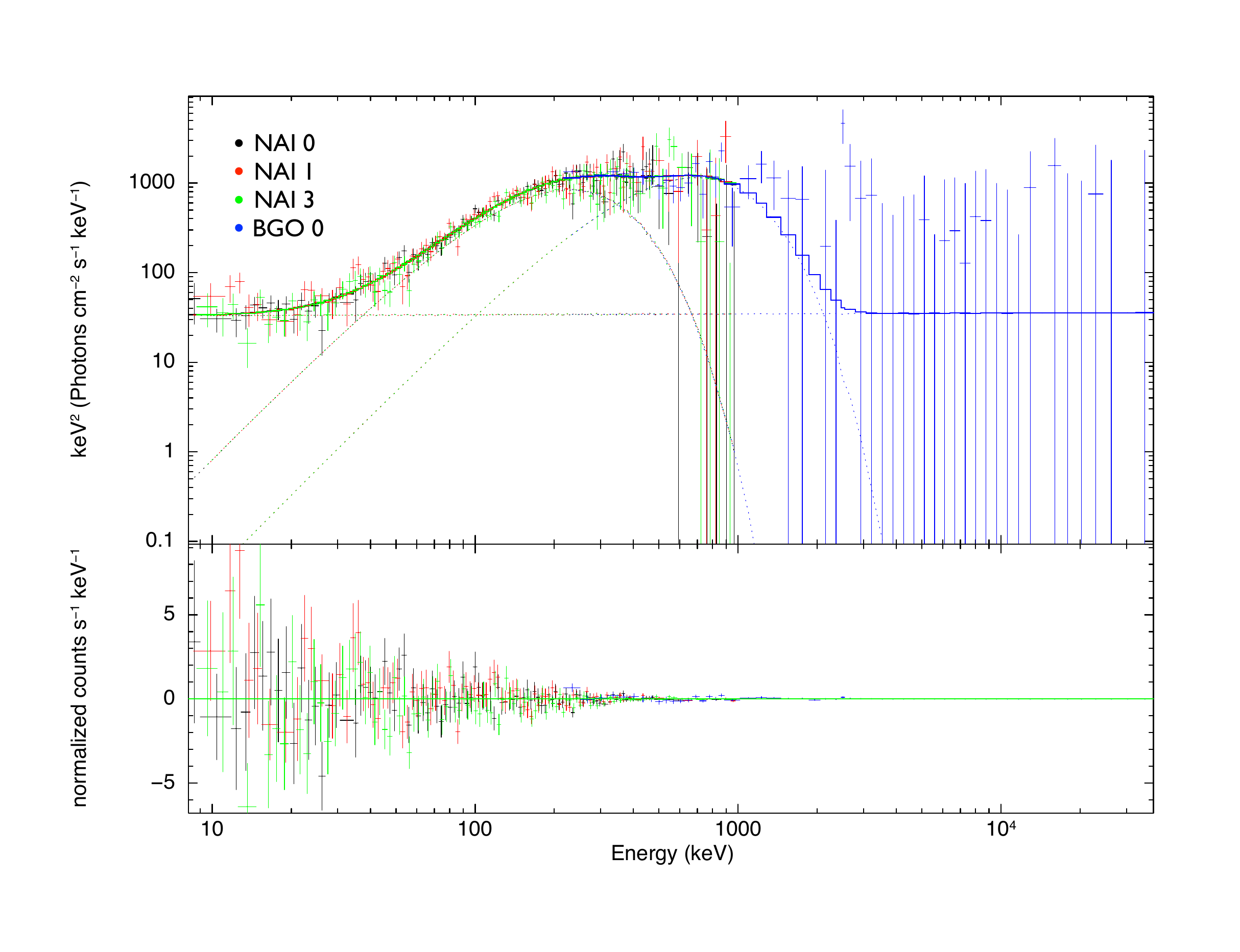}}
\caption{Spectral fits to the time {interval} $11.28$ s $- 15.2 $ s, green/solid line for NaI detectors and blue/solid line for BGO detector. (a) Band function fits.
The residuals are wavy close to the peak of the spectrum and it clearly does not fit the data at lower energies. (b) Band + BB fit, the wavy structure of the residuals is less pronounced, but still remains. (c) The best fit model, Comptonised (2 BB) + power law.}
\label{fig:spectra_pl}
\end{center}
\end{figure*}

A sample spectral fit for GRB110920A  is shown in the upper left panel in Figure \ref{fig:spectra_pl}. For an acceptable fit the residuals are expected to be random.  It is thus obvious from the wavy structure of the residuals  that the Band function alone does not reproduce the actual spectral distribution of the data.  This fact strongly suggest that the spectrum is more complicated than a Band function alone. Additional spectral breaks can be imagined as well as additional spectral components.

The main problem with these fits is the inability of the low-energy power-law to capture the additional curvature, while the peak energy is well determined. The Band function fits do, however,  indicate that the spectra are exceptionally hard and narrow (see also \S \ref{sec:SW}). 
The determined photon index, $\alpha$, is found to lie above -0.4 through out the burst, which is inconsistent with both slow cooling synchrotron ($\alpha \le$ -0.67)\footnote{The expected measured value for slow cooled synchrotron emission is even softer, closer to $\alpha \sim -0.8$ (Burgess et al. 2014)} as well as fast cooling synchrotron ($\alpha \le$ -1.5) emission processes. We also find that $\alpha$ increases linearly with time from $-0.4$ to $+0.8$. This is contrary to the typically observed behaviours where $\alpha$ evolves from hard to soft \citep{Crider1997, Kaneko2006, Ghirlanda2003}. In particular, after $70$ s, $\alpha$ lies above $0$ making it incompatible with any optically-thin, non-thermal processes and these spectra are among the hardest GRB spectra ever observed \citep{Ryde2010,Kaneko2006}. Moreover, we find that $E_{\rm peak}$ and $\alpha$ have a negative correlation: $E_{\rm peak}$ decays almost exponentially with respect to $\alpha$, which is clearly opposite to what is typically observed \citep{Kaneko2006}. The high-energy photon-index, $\beta$, is very hard in most bins and mainly has upper limits.

\begin{figure}[h]
\begin{center}
\resizebox{84mm}{!}{\includegraphics{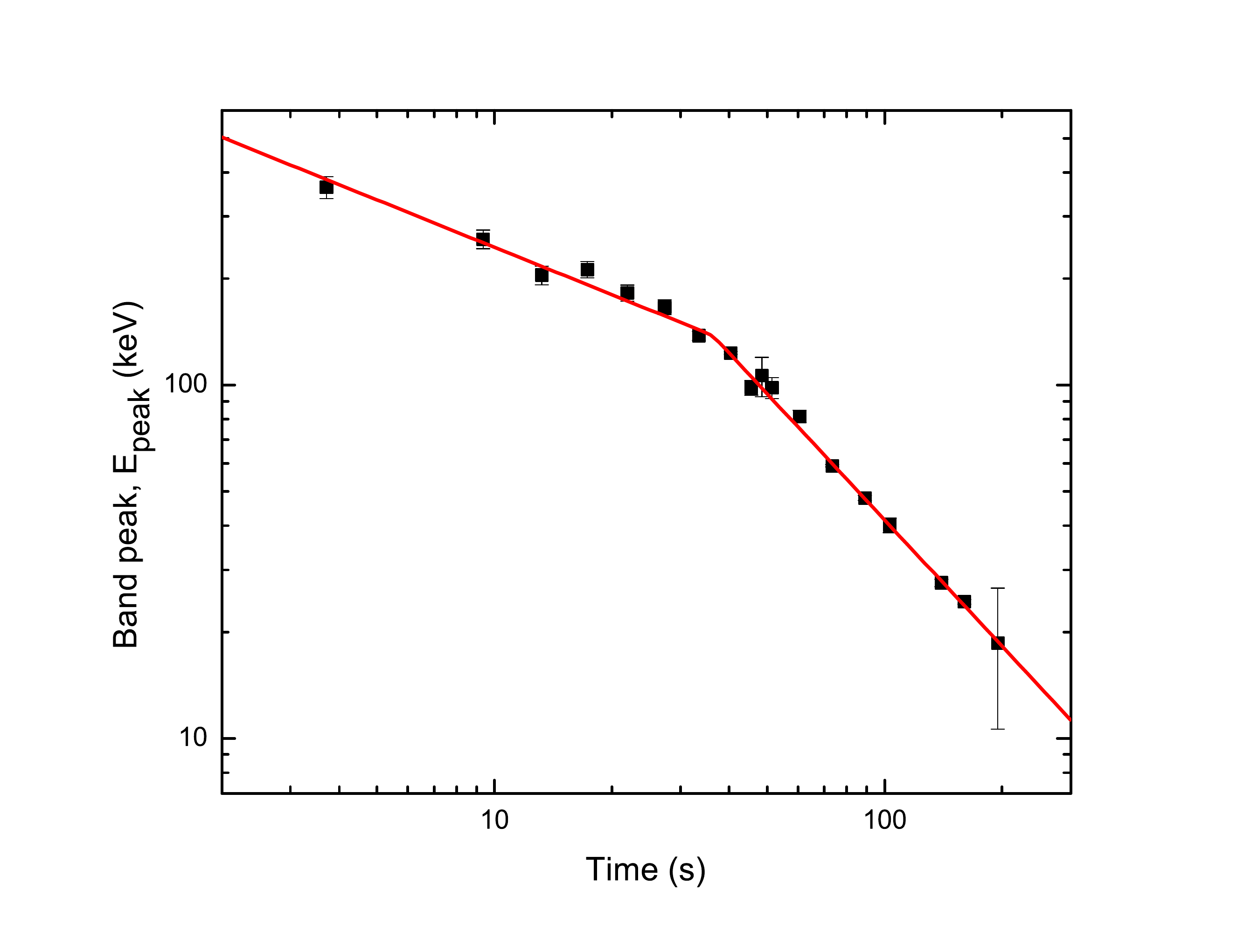}}
\caption{The Band function peak, $E_{\rm peak}$,  evolves as a broken power law from $700$ keV to $50$ keV with the break occurring at $\sim 28$ s. Note that these fits are for the Band only model, which is not the best fit model.}
\label{fig:bandonly}
\end{center}
\end{figure}

Figure \ref{fig:bandonly}   shows that the Band $E_{\rm peak}$ decreases from nearly $700$ keV to $50$ keV, following a broken power law with a break at $\sim 28 \pm 3 $ s. We note that this break does not  coincide with the flux break at $\sim 12$ s. The temporal power-law index before (after) the break is $-0.35 \pm 0.07$ ($-0.95 \pm 0.02$). 

\subsubsection{Band + blackbody fits}
\label{sec:bandbb}
Multicomponent spectra for GRBs were first suggested by \cite{Meszaros&Rees2000}, where, e.g.,  a blackbody models the photosphere and a non-thermal emission is expected from an optically-thin emission process (two emission-zone model, see also \cite{Meszaros2002,Zhang&Meszaros2002,Daigne&Mochkovitch2002}. Indeed,  
\cite{Ryde2004, Ryde2005} successfully fitted such a model to data from the {\it CGRO} Burst and Transient Source Experiment (BATSE). Due to the limited energy range of the detector ($\sim 20 -2000 $ keV), the model was limited to a blackbody  and a power-law function modelling the non-thermal emission. Similarly, {\it Fermi} observations ($8$ keV -- $> 40$ MeV) have been analysed with such a model in mind, where a blackbody is instead combined with a Band function  (e.g., \citet{Guiriec2011}, \cite{Axelsson2012}), or synchrotron emission \citep{Burgess2014a,Yu2015}.
 The Band (or synchrotron) model adds a second spectral peak at higher energies, producing  double humped spectra. 

\begin{figure}
\begin{center}
\resizebox{84mm}{!}{\includegraphics{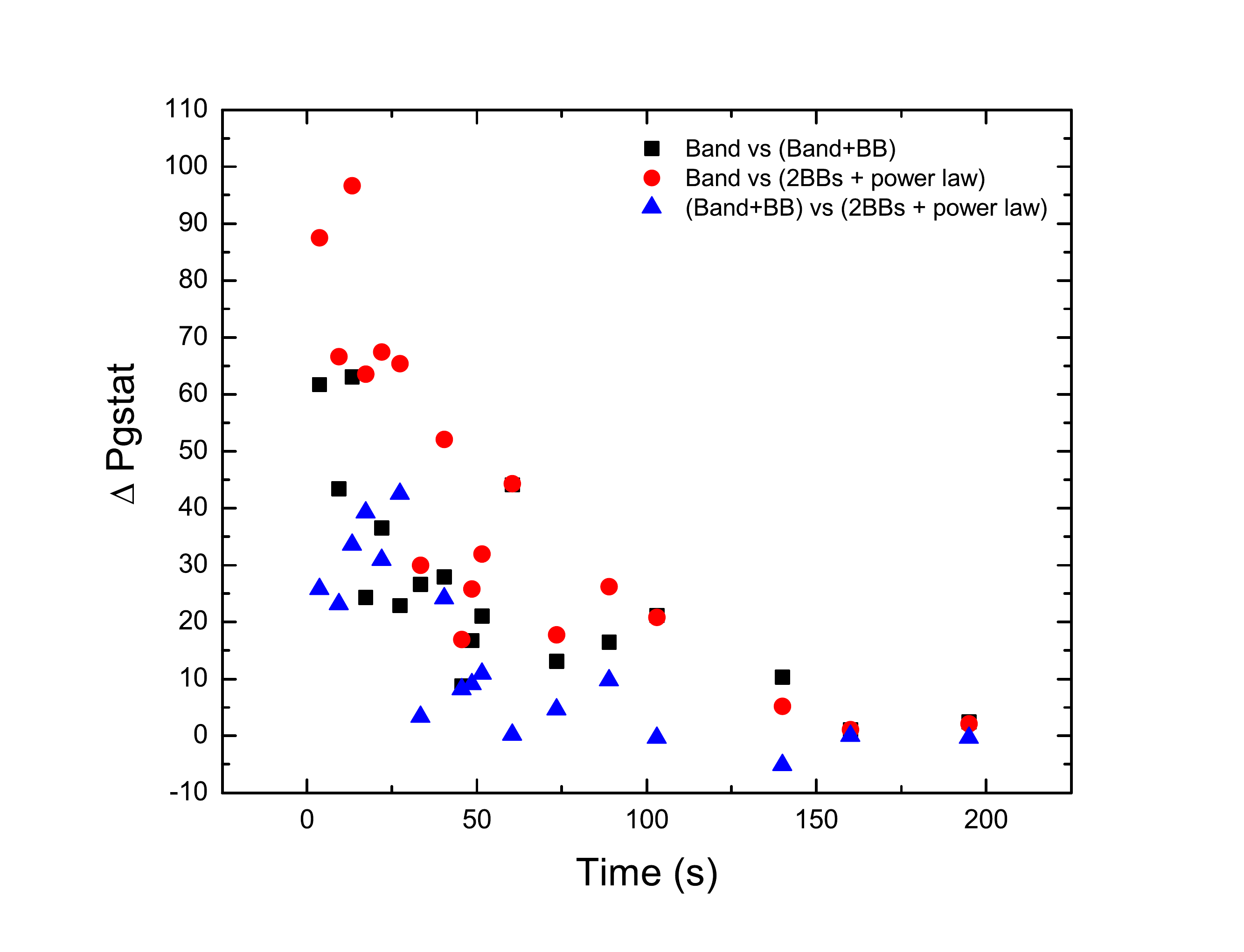}}
\caption{The difference in pgstat for different models (Band+ BB (black/squares) and Comptonisation (2BBs)+ power law (red/circle)) with respect to the Band function fits done to the spectrum. Also, the difference in pgstat of the model Comptonisation (2BBs)+ power law with respect to Band + BB model is shown (blue/triangle). }
\label{fig:delta_pgstat}
\end{center}
\end{figure}

Thus, using the above motivation, a combination of Band function and blackbody was fitted to the spectra of the individual time bins through out  the burst (see also \cite{McGlynn2012, Shenoy2013, Burgess2014a}. This model is a better fit with respect to the Band function model as the residuals become more random (upper right panel in Figure \ref{fig:spectra_pl}).  Moreover, the fits significantly improve the pgstat\footnote{{\tiny http://heasarc.gsfc.nasa.gov/xanadu/xspec/manual/XSappendixStatistics}} by $> 25$ in certain time bins.  In Figure {\ref{fig:delta_pgstat}} we show the difference in pgstat, with respect to the Band function fits, for different models used for the analysis.

\begin{figure*}
\begin{center}
\resizebox{84mm}{!}{\includegraphics{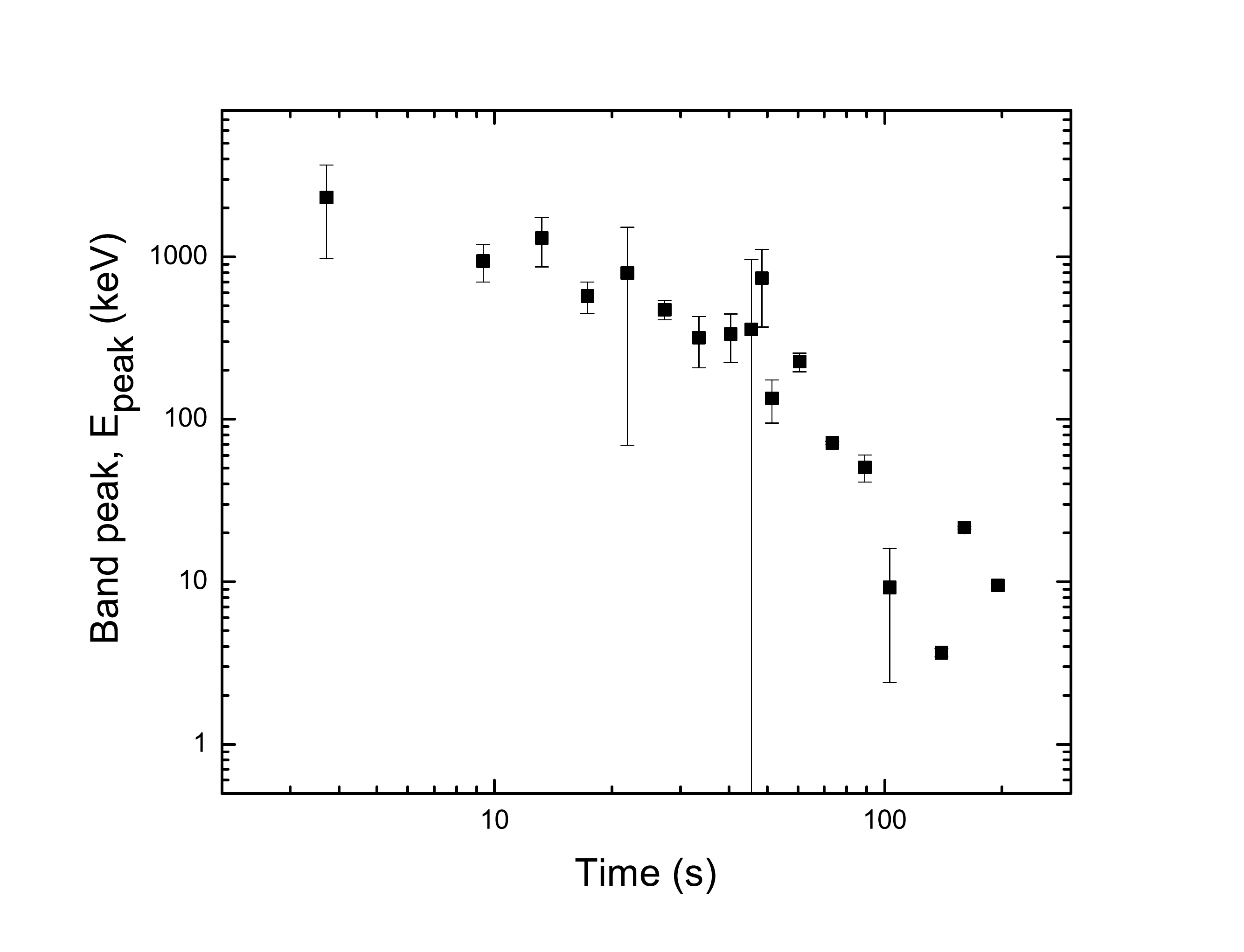}}
\resizebox{84mm}{!}{\includegraphics{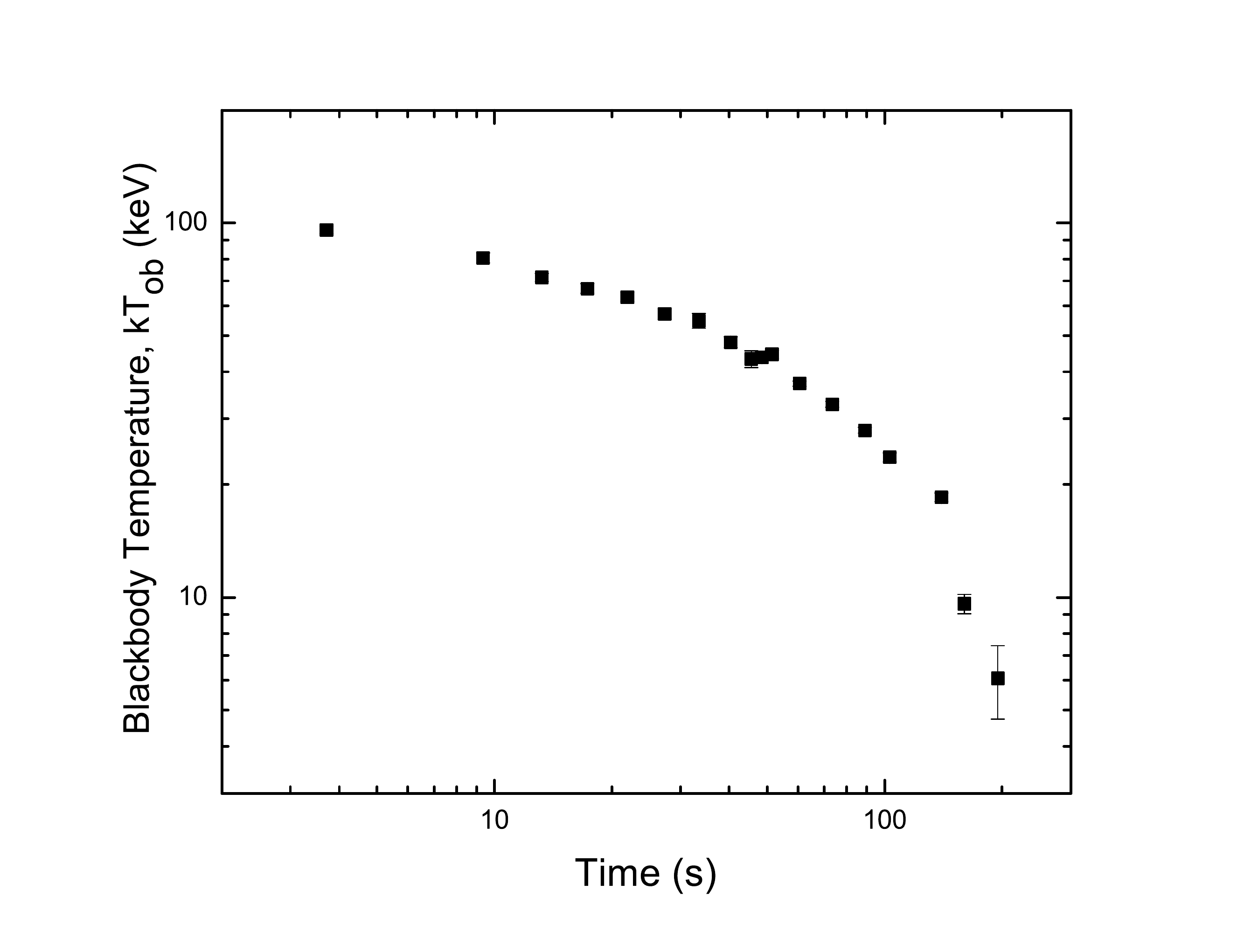}}
\caption{Results of Band function + blackbody fits: (a) The Band peak, $E_{\rm peak}$,  decreases
from $2$ MeV to $20$ keV with no significant break.  
(b) The blackbody temperature, $kT$ decreases smoothly and monotonously with time, in contradiction to the typical broken power-law observed. }
\label{fig:bandbb}
\end{center}
\end{figure*}

 The fit results (Band $E_{\rm peak}$ and k$T$) are shown in Figure \ref{fig:bandbb}. When a blackbody is added to the Band function, the blackbody captures the low-energy spectral break and the Band $E_{\rm peak}$ gets pushed to higher values compared to the peak values found in the  Band-only fits. This results in  the  residuals becoming more random which implies a better fit to the data. The presence of a blackbody leads to the low-energy photon-index, $\alpha$, becoming softer, as is generally observed (see, e.g., \citet{Guiriec2013}, but see also \cite{Burgess2014a}). During the burst, $E_{\rm peak}$ is observed to decrease from nearly 2 MeV down to approximately 20 keV. The Band function narrows with time and towards the end of the burst, the spectrum around the $\nu F_{\nu}$ peak becomes nearly as narrow as a blackbody.  
The low-energy power-law index $\alpha$ increases  from -1 to  values larger than 1. It is worth noting that in this model also, after $100$ s, $\alpha$ becomes harder than 0. However,  we note that before $100$ s, the value is nearly constant at  $\alpha \sim -1$ which is consistent with slow cooling  and (modified) fast cooling synchrotron emission. The high-energy photon-index, $\beta$, is very hard in most bins and has only upper limits. 
The correlation between the  Band $E_{\rm peak}$ and $\alpha$ does not show a clear negative correlation, since $\alpha$ is nearly constant until $100$ s. 

The blackbody temperature, $kT$, decreases smoothly with time with no significant break (Fig. \ref{fig:bandbb}). This is  in contrast to what is typically observed where the temperature decay has a characteristic temporal break \citep{Ryde2004, Ryde&Pe'er2009, Axelsson2012, Penacchioni2012, Burgess2014a}. The relative strength of the non-thermal component decreases with time: ratio of the blackbody flux to the total observed flux, $F_{\rm BB}/F_{\rm tot}$, evolves from $\sim 30 \%$ to nearly $90 \%$, which is larger than typically observed values \citep{Burgess2014a}.

Alternative fits for a two-zone model for GRB110920A have been performed by \cite{Burgess2014a,Yu2015}, who replaced the Band function with an optically-thin synchrotron emission function. They find that only slow cooled synchrotron emission is allowed by the data, mainly due to the restrictive curvature of the spectrum around its peak.

\subsection{Best fit model: {\bf 2 blackbodies + power law}}
\label{bb2pow}
As obvious from the above fits (\S \ref{sec:band}), the spectra are more complex than a simple Band function. The shape of the spectra near the spectral peak is more of a top-hat; at the same time the spectra are very narrow. In addition to this, the low-energy break in the spectrum cannot be well captured by neither  the Band function nor the BB + Band function fits. This motivates us to explore a Comptonisation model. Indeed, in \cite{Peer&Waxman2004} and Pe'er et al. 2006, it was shown that the process of Comptonisation can result in such top-hat spectral shapes (see also discussion in \cite{Ghisellini1999}).

\subsubsection{Motivation of empirical model}

 Assume that the photospheric emission, that is approximated by a blackbody with temperature $T_{\rm i}$, undergoes 
Comptonisation by  a thermalised pool of energetic electrons of temperature $T_{\rm e}$. The resulting spectrum will then be 
characterised by two temperatures  $T_{\rm i}$ and  $T_{\rm e}$. Under certain circumstances the main features of such a spectrum can be captured by an approximation consisting of  two blackbodies, with temperatures $T_{\rm i}$ and $T_{\rm e}$.  We will show in Iyyani et al. (2015 in prep.) that such an approximation is  valid for ratios of $T_{\rm e}/T_{\rm i}$ not much larger than $\sim 10$, optical depth less than a few tens and when there is no significant synchrotron photon production as a result of dissipation. 
 Furthermore, an underlying assumption of using the high-energy blackbody is that the electron distribution is Maxwellian. This is, in general, a good approximation for radiation mediated shocks below the photosphere (see, e.g., \cite{Bromberg2011}).

The advantage of using a simple analytical model is that  the fitting routine becomes 
greatly simplified compared to a fitting routine invoking a fully developed Comptonisation model (see Ahlgren et al. 2015, in prep.) for the fits, but still retaining the important features. This is in particular important when exploring such spectral behaviour in large amounts of data.

The two-blackbody approximation further neglects any broadening effects due to high-latitude emission (Pe'er 2008, Lundman et al. 2013). This is motivated by the fact that several observed burst spectra are inconsistent with the shape expected from the simplest models of such broadening. This is particular the case for the blackbody bursts in Ryde (2004),  and more recently for GRB100507  \cite{Ghirlanda2013} and GRB101219B  \cite{Larsson2015}.  These bursts point towards the emitting region not being spherically symmetric (which is typically the underlying assumption), but rather containing patches on scales smaller than $1/\Gamma$. Similarly small-patch emission zones are also found in the jet simulations \citep{Lopez-Camara2013}. In particular, as shown in  \S \ref{sec:SW},  the width for GRB110920A is also inconsistent with such broadening, in particular, towards the end of the burst. This argues that such effects should be secondary in GRB110920A.

In \S \ref{sec:cal_outflow} and \S \ref{sec:compt} we illustrate how the two-blackbody approximation is used in interpreting the Comptonisation process.

With  this motivation we use  a fitting function that combines two blackbodies, that captures the main Comptonised component, and a power-law function that, over the observed energy band, captures any secondary, broad-band emission. Such a power-law component was indeed found to accompany the photospheric emission component  in GRB090902B \citep{Ryde2010} and was suggested  to be due to
dissipation processes well above the photosphere  \citep{Pe'er&Ryde2011}. We note further that the fit function consisting of two blackbodies and a power-law function has equal number of parameters as the Band + blackbody model, which simplifies model comparisons. 

\subsubsection{Results of the empirical fits}

The model fits the data very well as can be seen in the example in Figure \ref{fig:spectra_pl}. We find that until $100$ s, this model is indeed a better fit with respect to the Band + BB fits and significantly improves the pgstat (by $> 35$ in certain time bins), see Figure {\ref{fig:delta_pgstat}}.
After $100$ s, we find the power-law component is no longer statistically needed. Therefore, the spectra after $100$ s are fitted only using two blackbodies, having merely four free parameters. 

\begin{figure*}
\begin{center}
\resizebox{130mm}{!}{\includegraphics{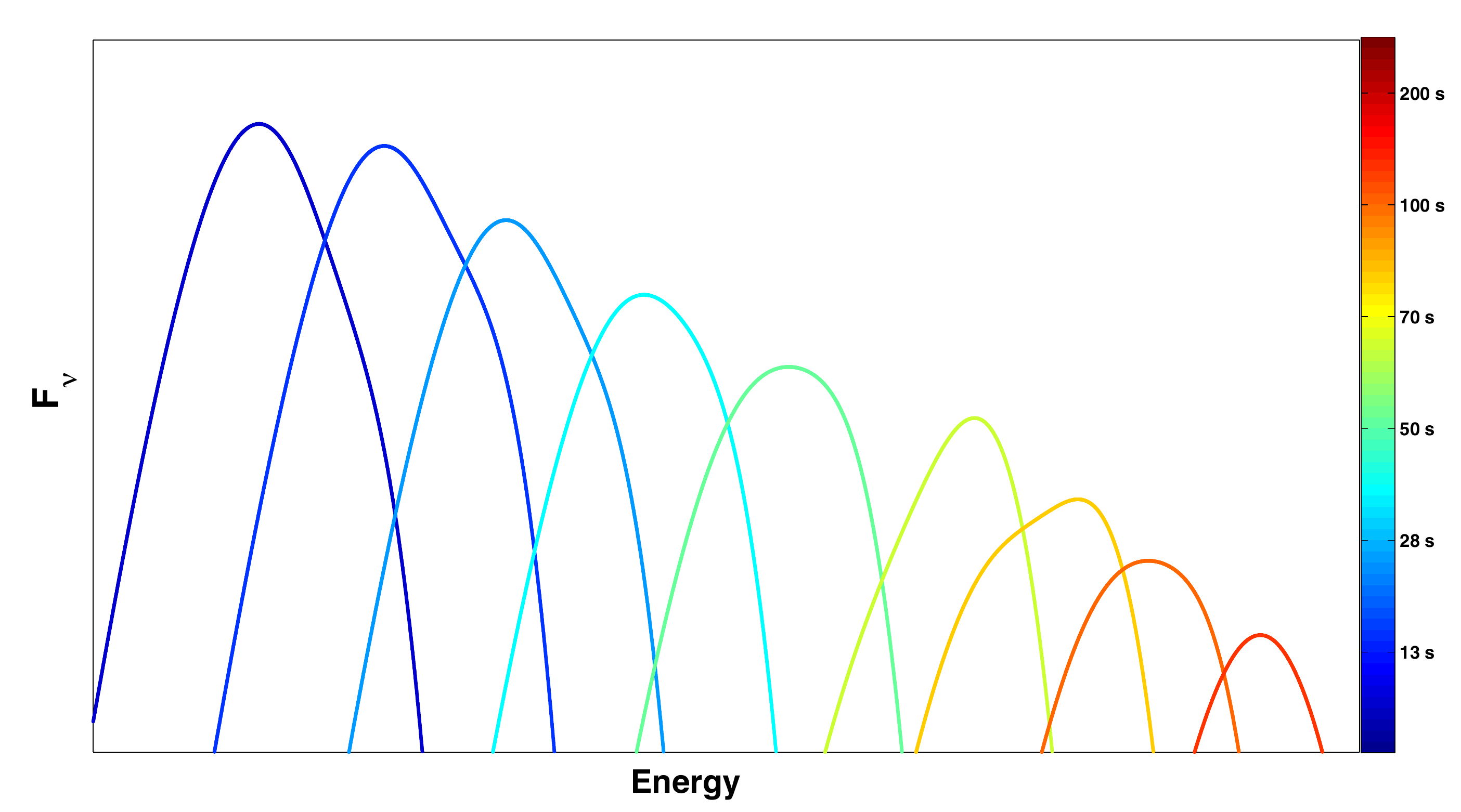}}
\caption{The evolution of the main spectral emission component of every second time bin is shown ($F_{\nu}$ versus energy).  
Note that the time-resolved spectra are arbitrarily shifted in energy in order to highlight the change in  spectral shape.  }
\label{fig:pl_Fnu_models}
\end{center}
\end{figure*}

In Figure \ref{fig:pl_Fnu_models} we plot the evolution of the $F_{\nu}$ spectra of only the Comptonised component (omitting the power-law component). In the figure the spectra are shifted in energy in order  to highlight the evolution of the spectral shape.

Since the model describes the shape of the spectrum well and also proves to be statistically significantly better than Band + BB fits in the beginning of the burst, we consider this model to be the best model for the burst.

\subsubsection{Best fit parameters}
\label{best_fit_param}
\begin{figure*}
\begin{center}
\resizebox{84mm}{!}{\includegraphics{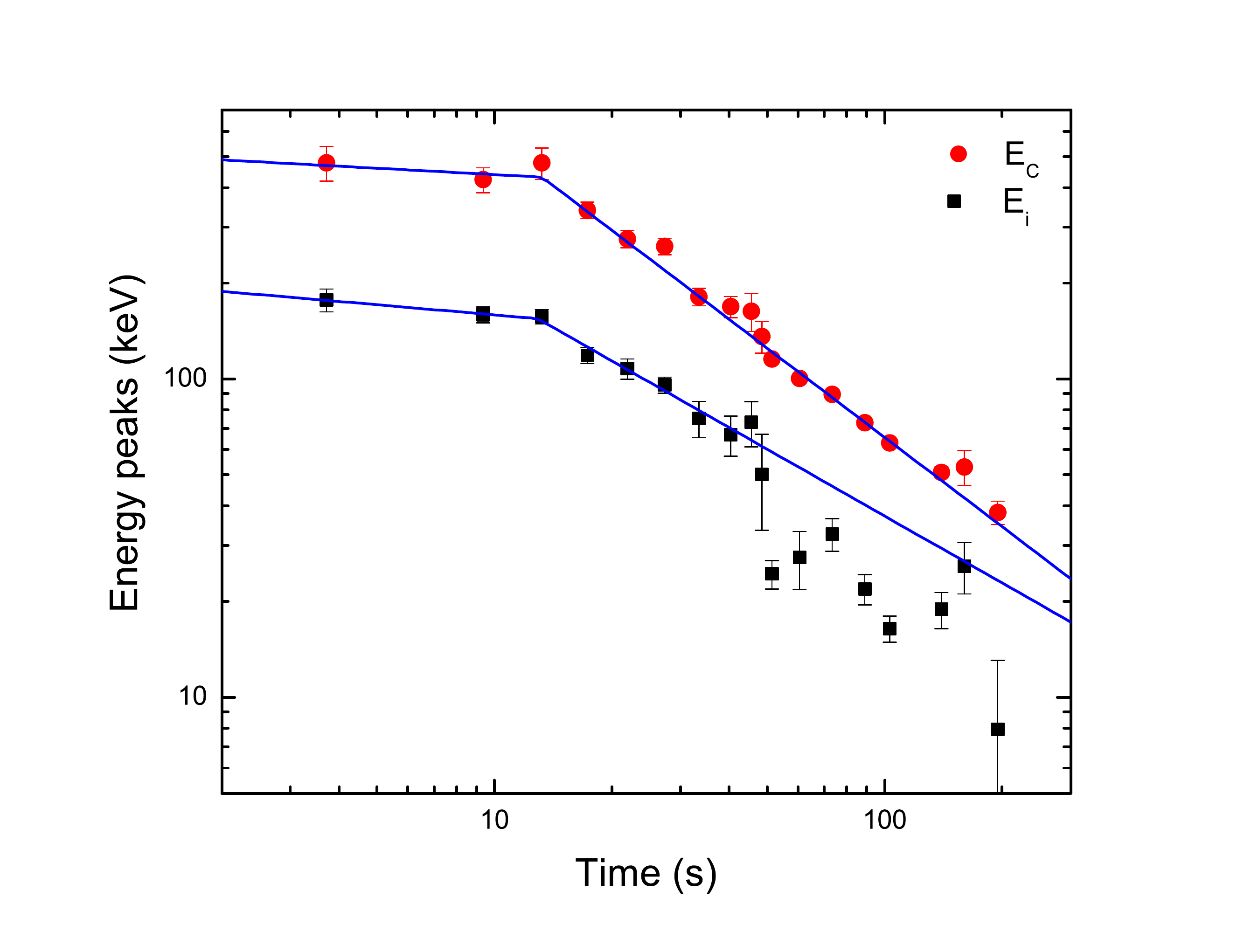}}
\resizebox{84mm}{!}{\includegraphics{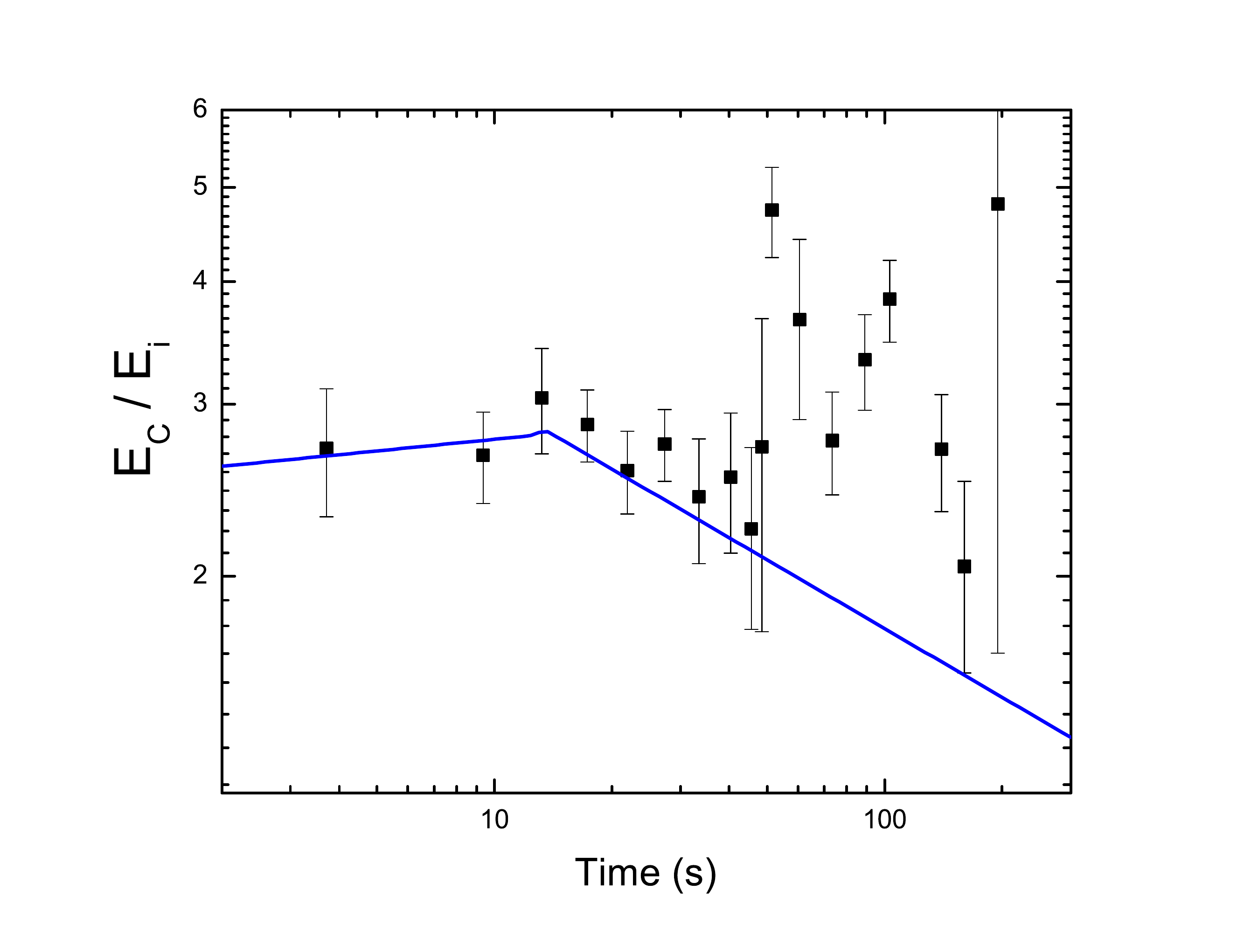}}
\resizebox{84mm}{!}{\includegraphics{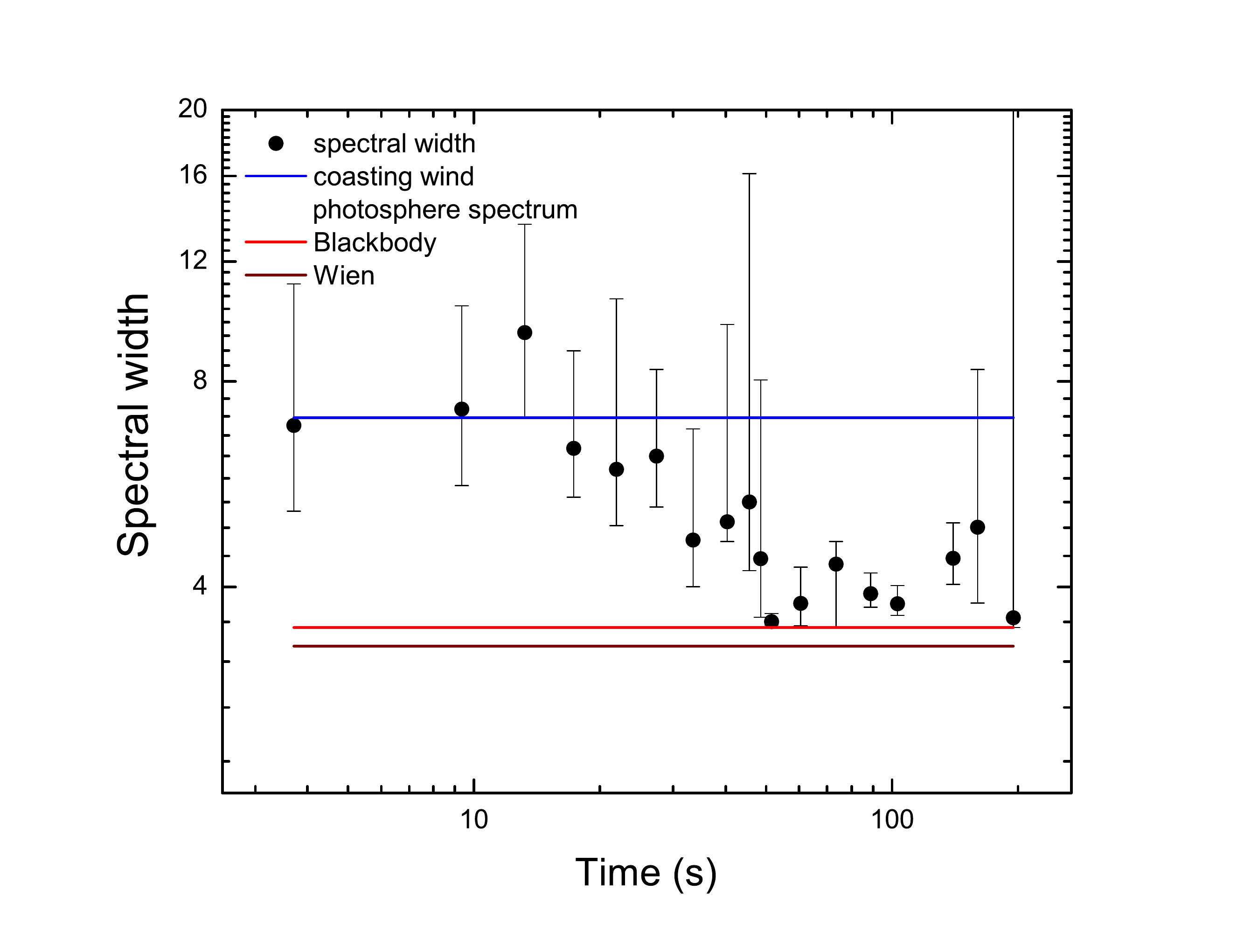}}
\resizebox{84mm}{!}{\includegraphics{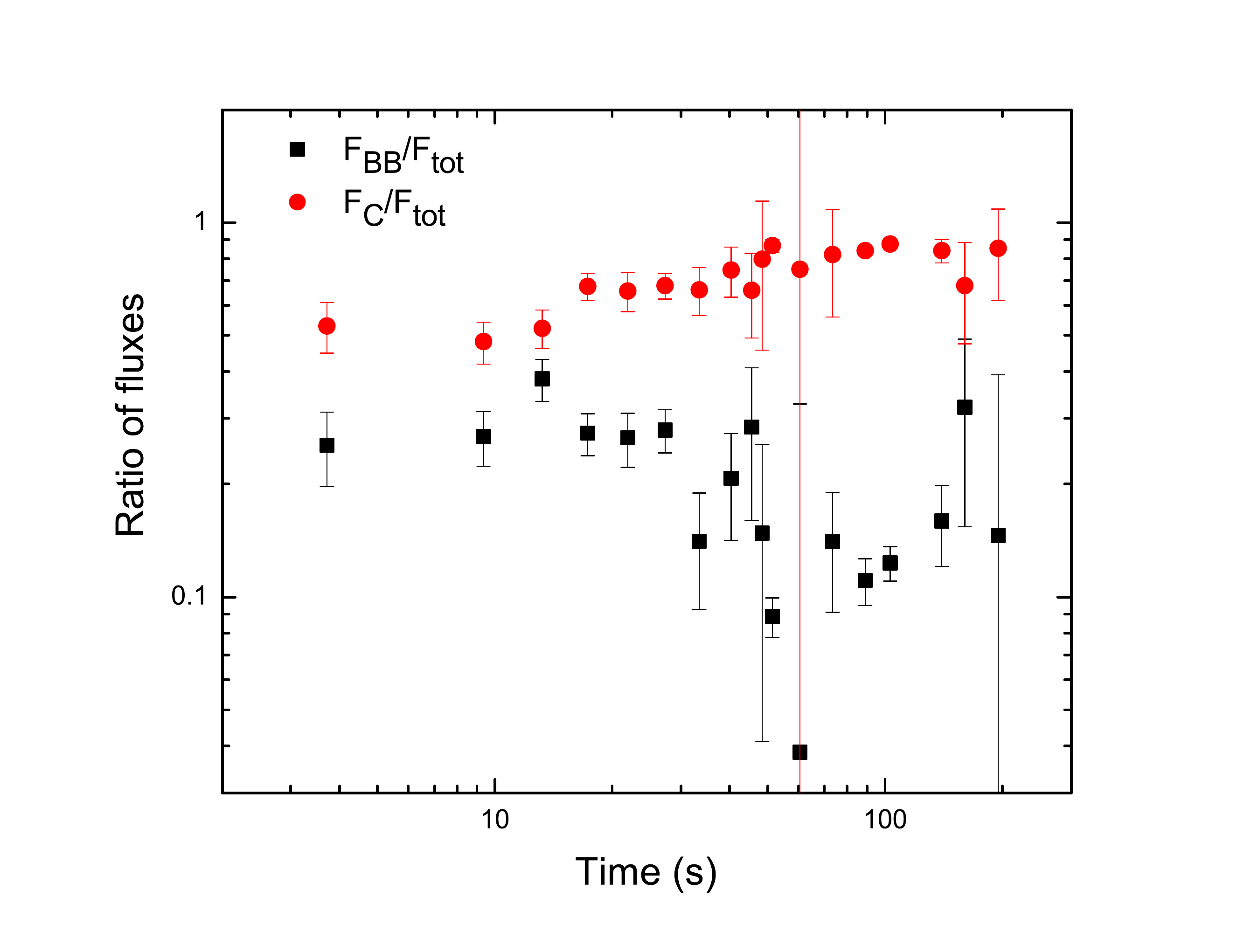}}
\caption{ Results of the best fit model (2BB + power law): (a) The peak energy of the low-energy peak, $E_{\rm i}$, and the high-energy peak, $E_{\rm C}$. The blue lines indicate the best fit to a broken power law function. Note the fit to $E_{\rm i}$ is only performed until $50$ s, after which it starts to deviate. (b) The ratio $E_{\rm C}$/$E_{\rm i}$; the solid line corresponds to the ratio of the best fit models in figure (a). 
(c) The spectral width of the $\nu F_{\nu}$ peak  decreases with time from $\sim 7$ to $3.6$,  approaching the width of a blackbody. (d) The fraction of the observed total flux residing in the low-energy BB ($F_{\rm BB}/F_{\rm tot}$) and  in the high-energy BB  ($F_{ \rm C}/F_{\rm tot}$). }
\label{fig:bb2pow}
\end{center}
\end{figure*}

Properties  of the fits are shown in Figure \ref{fig:bb2pow}.  
The peak of the low-energy blackbody (BB1) \footnote{Note that the thermal peak is given by $E_{\rm  i} = 2.7 kT_{\rm i}$ where $T_{\rm i}$ is the temperature of low-energy blackbody}, $E_{\rm i}$, decreases from nearly $200$ keV to $8$ keV with a break at $12.9 \pm 1.7 $ s (Fig. \ref{fig:bb2pow}a). The break coincides with the peak in the light-curve. The power-law indices are {before (after) the break $-0.1 \pm 0.08 (-0.69 \pm 0.06)$.} 

This evolution is very different from the cooling behaviour found in the Band+BB fits above, where the temperature did not show any break in its evolution. The break in the $E_{\rm i} $ evolution found here is, on the other hand, in agreement with what is typically observed for cooling blackbodies \citep{Ryde2004, Ryde&Pe'er2009, Penacchioni2012, Axelsson2012}. This fact is indeed reassuring and strengthens the conclusion that the correct spectral model is used.

The spectral peak of the high energy blackbody (BB2)\footnote{Note that the thermal peak is given by $E_{\rm C} = 2.7 kT_{\rm e}$ where $T_{\rm e}$ is the temperature of high-energy blackbody}, $E_{\rm C}$, also decreases as a broken power-law from nearly $500$ keV to $40$ keV, again with a break at $13.2 \pm 2.4$ s, which is consistent with the break in $E_{\rm i}$. The power-law indices are {before (after) the break $-0.06 \pm 0.19 (-0.93 \pm 0.03)$.  We note that the post-break index is significantly steeper than for $E_{\rm i} = E_{\rm i}(t)$.  This is also illustrated  by Figure \ref{fig:bb2pow}b, in which the ratio $E_{\rm C}/E_{\rm i}$ is plotted. The model-ratio is shown as a blue line, has a maximum of $\sim 3$ and gets smaller with time. This suggests a decreasing spacing between the peaks (see further the discussion in \S \ref{sec:cooling})}. Moreover, different from the evolution of $E_{\rm C}$, after 50 s, $E_{\rm i}(t)$ does not show any clear trend, and does not follow the initial power-law decay,  and exhibits a fluctuating nature. Thus, $E_{\rm i}$ is highly correlated with $E_{\rm C}$ only until 50 s.

Apart from the main emission component,  the power-law component also varies, with its index changing from -1.5 to nearly -4, but it gets more poorly determined as the burst progresses. The power-law mainly accounts for the emission in the BGO as well as for the emission below $30$ keV. However, after $65$ s, there is no more detected BGO emission and after $100$ s, the emission below $30$ keV becomes less significant. The power-law component of the fits thus becomes insignificant and the spectra are consistent with only the Comptonised component (two blackbodies).  

The flux of the low-energy blackbody, $F_{\rm BB}$, is around $30\%$ of the total observed flux, $F_{\rm tot}$, while the flux of the high-energy blackbody
, $F_{C}$, dominates the total observed flux through out the burst (Fig. \ref{fig:bb2pow}d).

Finally,  we note that the timescale of variations observed in the light-curve of the burst is much longer than the dynamical time which for the typical values of photospheric radius, $r_{\rm ph} = 10^{12}$ cm and Lorentz factor, $\Gamma = 300$, is given by $r_{\rm ph}/(2 \Gamma^2 c) \sim 0.2 $ ms. This tells us that the central engine of the burst varies over a longer timescale and that the flow can be regarded as quasi-static in time. This validates the assumption that every time-bin can therefore be analysed assuming a steady flow.

\subsection{Spectral Width}
\label{sec:SW}
 
Different emission mechanisms produce predictable  widths of their emission spectra.  The spectral width, $SW$, is here given by the ratio of the two extreme values of energy at which the  $\nu F_{\nu}$ value of the spectrum is equal to half its maximum peak \citep{Axelsson2015}. 
For instance, a blackbody function is very narrow $SW = 3.5$,  while the observed spectrum from a non-dissipative photosphere, assuming a spherically symmetric, coasting wind has SW = 7 \citep{Lundman2013}. This is since the spectrum is a multicolour blackbody, rather than a pure blackbody due to additional flux of photons emitted at high latitudes\footnote{High-latitude emission is here defined as the emission that originates from angles larger than $1/\Gamma$ to the line-of-sight of the observer.}.  In comparison, synchrotron spectra are even broader. The spectral width of synchrotron emission from a mono-energetic electron distribution is $SW = 8.6$. However, such a distribution is not generally expected.  Fast cooling synchrotron emission (power-law electron distribution) has  $SW \sim 370$, while slow cooling synchrotron emission has $SW = 145$ (assuming the electrons distribution is a combination of Maxwellian + power law \citep{Baring1995, Baring2011, Summerlin2012}. 

We, therefore, measured the spectral widths from the fits. The spectral width of the Band-only fits decreases with time from nearly $SW = 7$ to  $4$. This illustrates  that the spectrum around the $\nu F_{\nu}$ peak is very narrow; the $SW$ even approaches the narrowness of a blackbody function. Throughout the entire duration of the burst, the $SW$ is smaller than what is typically observed: the distribution of observed peak spectral widths has a maximum at $ SW  \sim 10$ (Axelsson et al. 2014, MNRAS, submitted).  

\begin{figure}
\begin{center}
\resizebox{84mm}{!}{\includegraphics{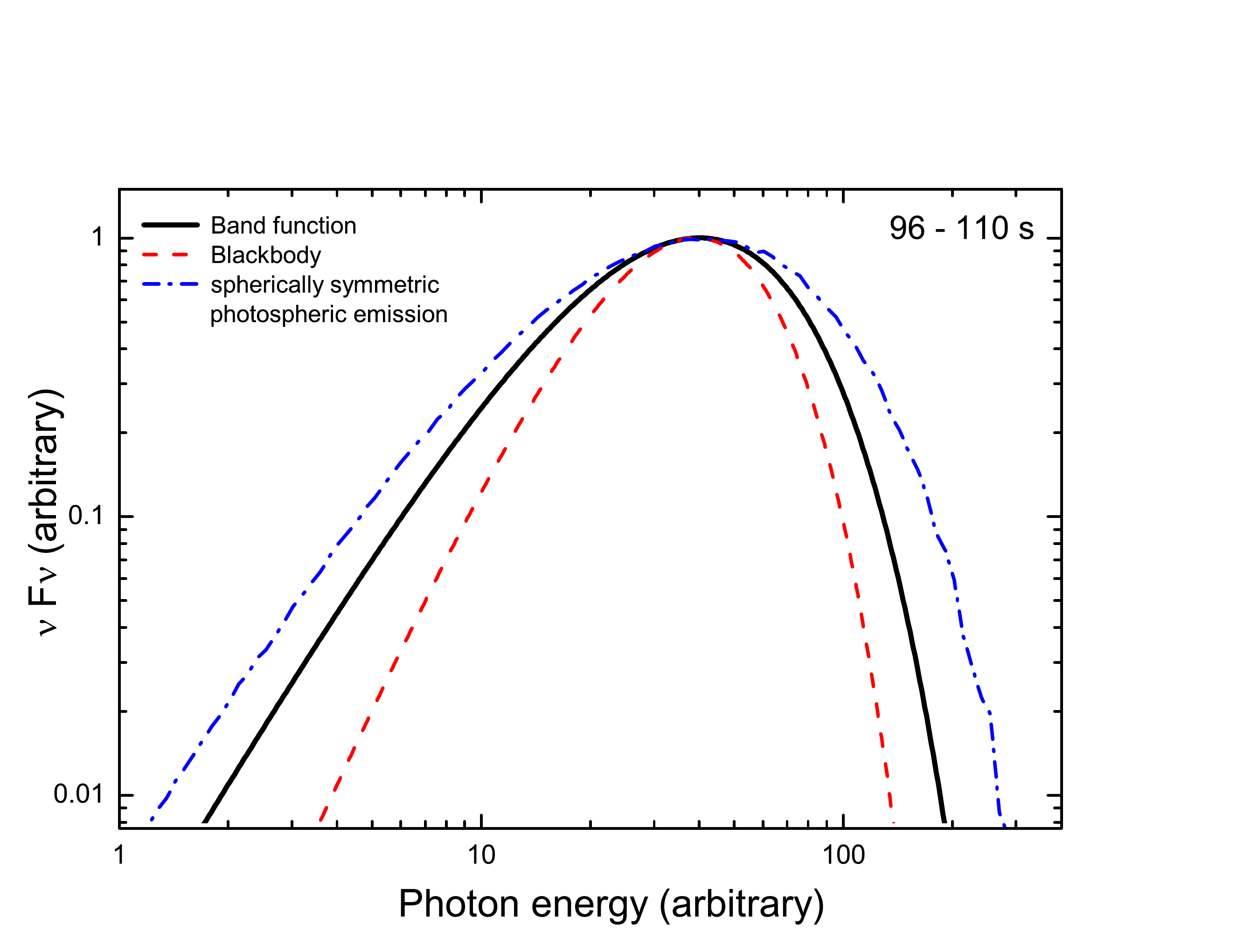}}
\caption{Narrowness of the spectrum. The Band function fit for the time interval $96$ s $-110$ s is shown in black (solid line). The blackbody (red/dash line) as well as the photospheric emission from a spherically symmetric outflow at the coasting phase (blue/dash-dot line) is also plotted for the same peak energy to show the narrowness of the spectrum. 
The spectral width of the blackbody is $SW= 3.5$, the photospheric emission from a spherically symmetric outflow is $SW=7$ and that of the Band function in this time bin is $5$. }
\label{fig:NarrowBand}
\end{center}
\end{figure}

As an example, the Band function fit for the time interval  $96 - 110$ s is  plotted  in Figure \ref{fig:NarrowBand}, together with a blackbody as well as the emission from a non-dissipative photosphere in a spherically symmetric outflow (broadened blackbody), aligned at the same $\nu F_{\nu}$ peak energy. The spectral width of the Band function fit  is  $SW = 5$. Finally, we also studied the spectral width of the best fit model (Comptonisation model, \S \ref{bb2pow}): The $SW$ decreases with time from 7 to 3.6 which is very  similar to what was observed in the case of Band-only fits (Fig. \ref{fig:bb2pow}c).

\section{Physical Scenario}
\label{Phys_scenario}

\subsection{Quasi-thermal Comptonisation}

As the best-fit spectra suggest, Comptonisation of a thermal photon component is a strong candidate in explaining the observed shapes. 
Quasi-thermal Comptonisation  in various scenarios has previously been suggested to explain GRB  spectra \citep{Liang1997a, Liang1997b, Ghisellini1999}. 

\cite{Rees&Meszaros2005} suggested that dissipation due to internal shocks below the photosphere can lead to Comptonisation of the thermal distribution of photons entrained in the outflow.  Such dissipation can also be due to hadronic collision shocks \citep{Beloborodov2011}, or recollimation shocks occurring due to interaction between the jet and the star as the jet traverses the envelope of the progenitor star \citep{Lopez-Camara2013, Duffell2014}. In particular, as the jet is ejected out of the star such shocks are expected \citep{Mizuta2013}.

At large optical depths, $\tau$, due to the balance between the heating (direct Compton scattering) and cooling (inverse Compton scattering), the electrons attain a steady state and remain sub-relativistic, with values of $\gamma \beta = 0.1 - 0.3$ \citep{Pe'er2005}.   The thermal photons that are entrained in the flow from the central engine will serve as seed photons that will be Compton scattered by the energetic electrons. Before entering the dissipation site, these photons have been fully thermalised deep in the flow, either at the central engine itself or at a larger distance at which the flow starts to accelerate (the jet nozzle), denoted here by $r_0$.  No magnetic field is therefore needed to form the seed photons.  
In such a case, when the seed thermal photons scatter off the sub-relativistic electrons, the change in energy per scattering of the photons is small.  At the same time, due to large number of scatterings, $n_{\rm sc}$,  the photons diffuse from the thermal pool to higher energies and eventually form a new peak as a result of saturated Comptonisation. 

The spectrum that is formed through Comptonisation, at such a dissipation site, will be advected with the flow until it reaches the photosphere and the photons are emitted to the observer. This includes adiabatic cooling of the photon field.

\subsection{Radial distribution of heating}
\label{sec:radial}

The spectral shape of the emission released at the photosphere in such a scenario will depend on the properties of the subphotospheric dissipation. Apart from the amount of kinetic energy that is converted, it will also depend on its properties such as the dissipation's temporal variation  and radial extent and its variation. For instance, even for steady flows, it can be envisioned that either the dissipation occurs at all radii with some prescribed radial dependence (continuous dissipation) or that the dissipation is localised at a certain radial position (localised dissipation).
Continuous dissipation can be due to radial and/or oblique shocks \citep{Rees&Meszaros2005, Peer2006, Lazzati2009}, or magnetic reconnection \citep{Giannios2008, Begue&Peer2014} and in general leads to the formation of smooth Band-like spectra. All detailed spectral features produced by a dissipation episode at a certain depth will, by necessity, be washed out by subsequent dissipations closer to the photosphere, where the jet properties have changed. Moreover, in the case of continuous dissipation, since efficient photon production is only expected deep down in the flow, the number-density of photons  will not be able to maintain the level required for full thermalisation. The resulting peak (at high optical depths) will thus be a Wien peak with a temperature higher than  a corresponding thermalised, blackbody spectrum.

However, in a non-dissipative flow, even without photon production, the spectrum remains thermalised and a blackbody distribution is retained. In such a case, a localised dissipation can distort the photon spectrum through Comptonisation of these blackbody photons
and complex spectral shapes are easily achieved (e.g. \cite{Pe'er2006}). Such localised dissipation can also be  due to radial and/or oblique shocks, collisional processes \citep{Beloborodov2013},  or due to some characteristic scale, e.g. the surface of the progenitor star \citep{Mizuta2013}.  

The existence of two distinct breaks in the spectrum of GRB110920A,  and the fact that the spectra are extremely narrow, therefore strongly favours a scenario where the dissipation occurs over a range for which the jet parameters do not vary significantly, for instance, over a relatively narrow range in radii (localised dissipation). The dissipation site can therefore be characterised by an effective optical depth or, correspondingly, a dissipation radius, $r_{\rm d}$. Note, however, that any effect of the dissipation would occur over a range of radii determined by the timescales of the interaction between the electrons and the photons. 

Furthermore, we argue that in the case of GRB110920A the dissipation radius, $r_{\rm d}$ lies above, or close to,  the saturation radius, $r_{\rm s}$. This is based on the fact that  the dissipation occurring in the flow alters the spectrum significantly.  There should thus be significant kinetic energy in the outflow that could be dissipated and given to electrons. However, if $r_{\rm d} < r_{\rm s}$, then the kinetic energy density of the outflow would be less than the photon energy density, and a significant deviation is not expected. Moreover, we do not expect to have dissipation during the acceleration phase.

\subsection{Seed photon component and outflow parameters} 
\label{sec:cal_outflow}

We have argued that the flow in GRB110920A is a non-dissipative, passive jet, apart from a local, major dissipation episode at $r_{\rm d}$. The low-energy break in the observed spectra can thus be identified  as the temperature of the seed photon distribution,  that is advected from the central engine, and  is injected in the dissipation site. This component  allows us to calculate the outflow properties following \cite{Pe'er2007}. 

However, many of  the seed photons must have been  scattered to higher energies due to the Comptonisation process. Therefore the normalisation of the fitted (low energy) blackbody does not correspond to that of the injected seed blackbody. The photon number is conserved in the scattering process, therefore the seed photon normalisation can be estimated from the total number of photons in the Comptonised photon distribution. We note further that, 
additional photons are not expected to be produced unless very deep in the flow \citep{Vurm2013} {or may be produced by synchrotron emission \citep{Beloborodov2013}, which however depends on the strength of the magnetic field which is considered to be very low here (see \S \ref{gen_scenario})}. 
Thus, the original number of photons in the fireball, $N_0$, is equal to the number of observed photons, $N_{BB1}+N_{BB2}$, where $N_{BB1}$ and  $N_{BB2}$ are the number of photons in the two fitted blackbody components respectively.  $N_{BB2}$ corresponds to  the number of photons scattered from the seed component.

Using the estimated  original number of photons we can calculate the Lorentz factor, $\Gamma$, the photosphere radius, $r_{\rm ph}$, the saturation radius $r_{\rm s}$, and the nozzle radius of the flow, $r_{\rm 0}$.  For these calculations to be valid we have to  assume that the fraction of kinetic energy dissipated is low (see \citet{Begue&Iyyani2014}) such that $\Gamma = \eta \equiv L_0/\dot M c^2$, where $L_0$ is the total burst energy and $\dot M$ is the mass ejection rate, see further \S \ref{sec:compt}. Furthermore, the estimation of these parameters depend on the radiative efficiency of the jet ($1/ Y$), which is unknown for GRB110920A. Here, $Y$ is the ratio of the total burst energy to the observed prompt $\gamma -$ ray emission.
 However, \cite{Racusin2011} estimated the radiative efficiency for a sample of 69 moderately bright bursts (with redshifts) and $Y$ was estimated to be around $2 \leq Y \leq 20$ with an approximate average value of  $Y\sim 5$.
In the following calculations we therefore use $Y=5$.

\begin{figure*}
\begin{center}
\resizebox{84mm}{!}{\includegraphics{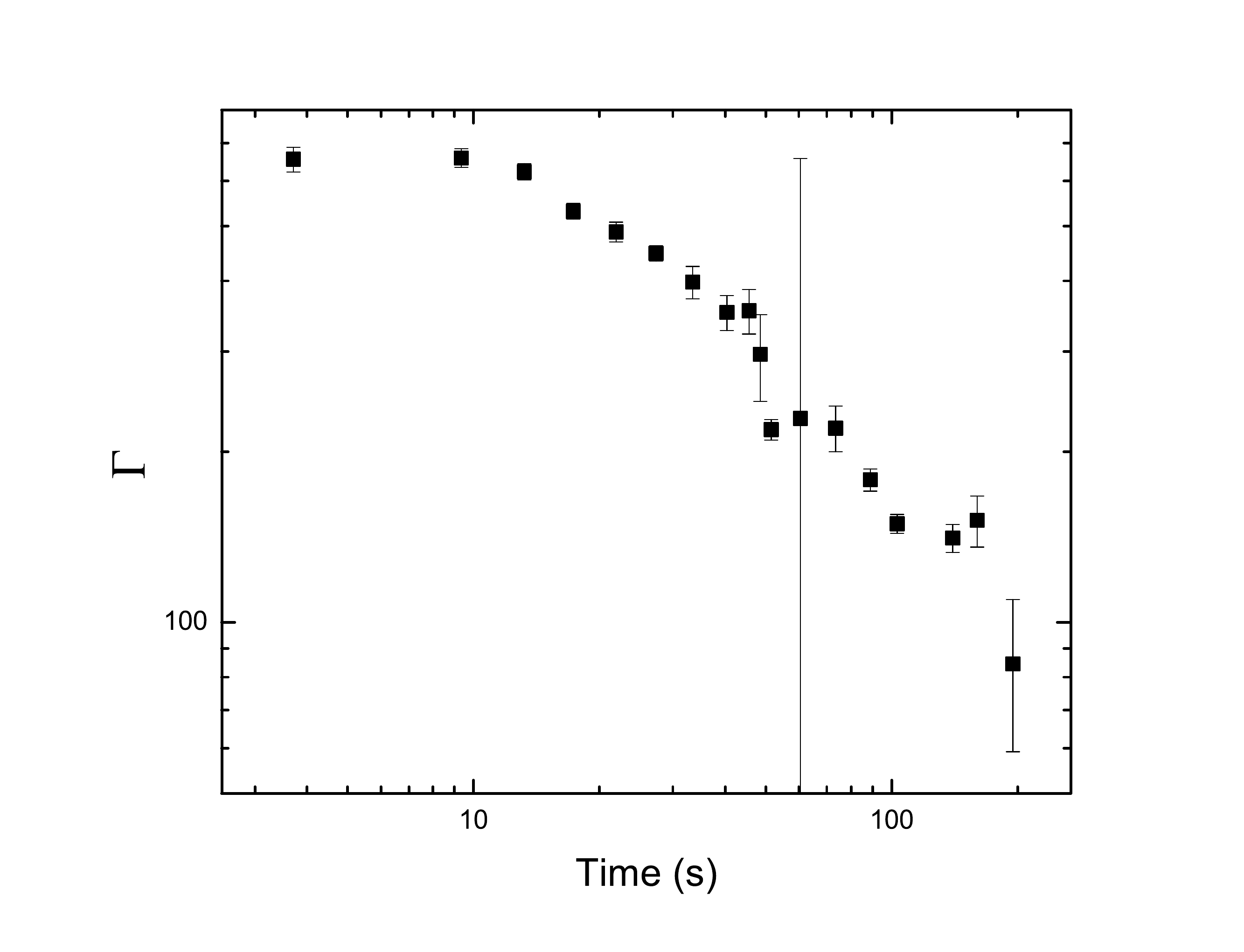}}
\resizebox{84mm}{!}{\includegraphics{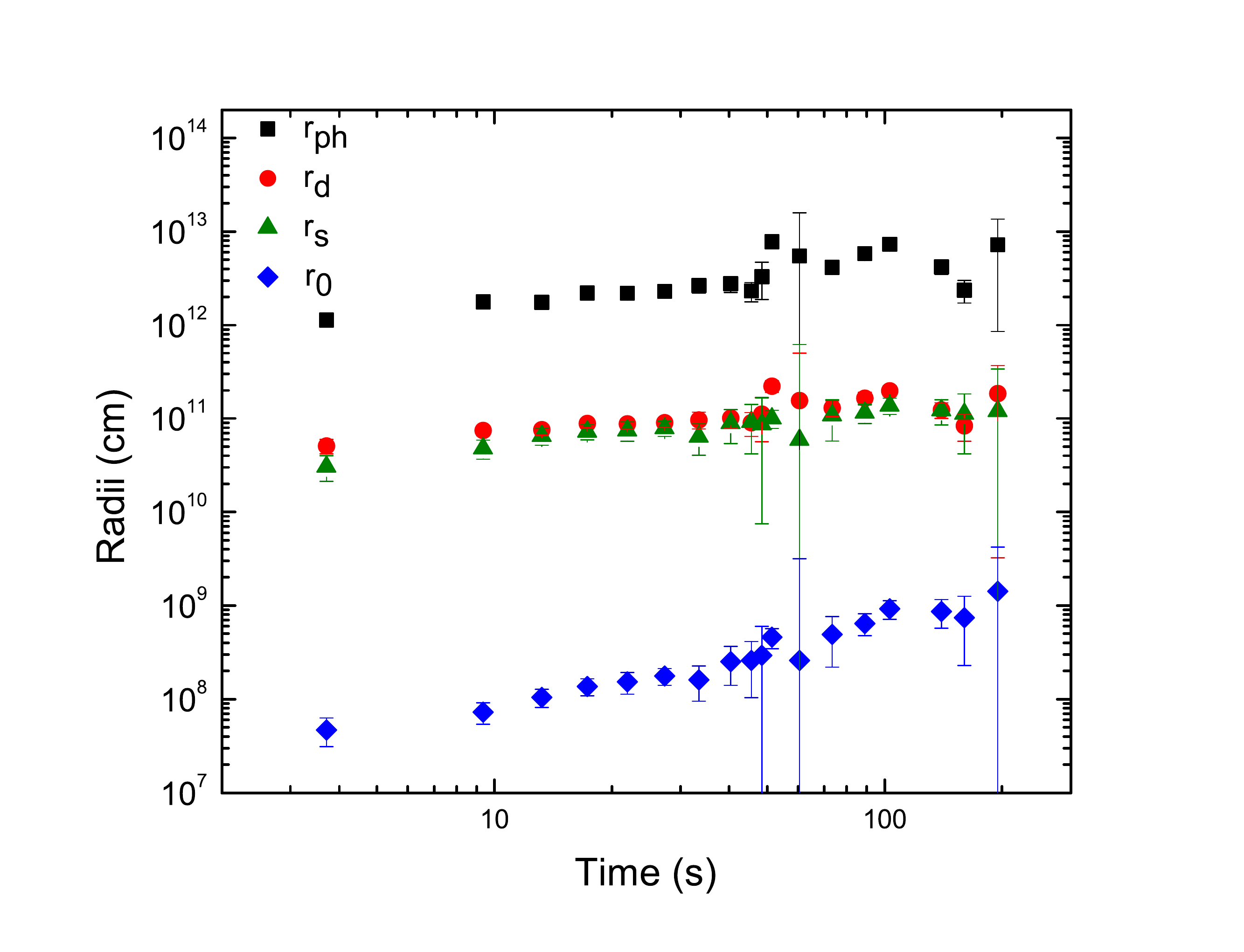}}
 
 \caption{ (a) The evolution of the Lorentz factor, $\Gamma$, of the outflow. (b) The evolution of the photospheric radius, $r_{\rm ph}$ (black/squares), the nozzle radius, $r_0$ (blue/diamond), the saturation radius, $r_{\rm s}$ (green/triangles). The evolution of the radius (upper limit) of the localised dissipation, $r_{\rm d, max}$ (red/circles).   All the parameters are estimated for a $Y = 5$ and a redshift, $z = 2$. }
\label{fig:outflow_par}
\end{center}
\end{figure*}

The estimated parameter values for a redshift, $z = 2$, are shown in Figure \ref{fig:outflow_par}a and \ref{fig:outflow_par}b.
For instance, for the time bin at 10 s, the values of the outflow parameters are:  $\Gamma = 660 \pm 25$, $r_{\rm ph} = 1.8 \pm 0.2 \times 10^{12}$ cm , $r_{\rm s} = 4.8 \pm 1.1 \times 10^{11}$ cm and $r_0 = 7.3 \pm 1.9 \times 10^{7}$ cm.  $\Gamma$ decreases as a broken power law, with a break at $ t = 15 \pm 2 $ s with a slope before (after) the break given by the power law index $-0.04 \pm 0.09\, (-0.71 \pm 0.04)$. The $r_{\rm ph}$ increases slightly until $50$ s ($r_{\rm ph} \propto t^{0.34 \pm 0.04}$) after which it does not show any particular trend. While $r_0$ increases as  $r_0 \propto t^{0.92 \pm 0.05}$,  $r_{\rm s}$ exhibits a weaker increase ($r_{\rm s} \propto t^{0.4 \pm 0.04}$). 
We note here that the estimated value of $r_0 \sim 10^8 -10^9$ cm (Fig. \ref{fig:outflow_par}b) lies within the range of the predicted values for the formation of the first collimation shock that is found in numerical simulations of hydrodynamical shocks moving though the envelope of the progenitor star \citep{Duffell2014}.  The existence of such a shock is a stable prediction in these simulations, and the value of $r_0$ is typically around $10^{9}$ cm, mainly depending on the dimensionless entropy of the flow. A high entropy flow (leading to high $\Gamma$) will have a smaller collimation shock radius and vice versa.

The above calculations assume a classical fireball model (Pe'er et al. 2007). We neglect, for instance,   effects of off-axis emission  (high-latitude effects) which have been suggested to be an explanation of the observed behavior of the temperature and flux \citep{Pe'er2008, Ryde&Pe'er2009}. Here, we instead assume that the varying observed properties  reflect varying properties of the flow at the central engine (dimensionless entropy, temperature, and luminosity) and radial (shock) dissipation pattern. For highly magnetised jets further alternatives exist by allowing the magnetisation at the central engine and magnetic dissipation pattern to vary \citep{Hascoet2013, Gao&Zhang2014}.

Finally, the outflow parameters are calculated for a redshift, $z = 2$, an averaged value of GRBs, since the redshift of the burst is unknown. We also assume a flat universe ($\Omega_{\lambda} = 0.73, H_0 = 71$). A different value of $z$ would only translate the values of the parameters depending on their dependences on $z$ (for e.g. $z = 0.3$ lowers the estimated values within a factor of 5) but will not change their temporal behaviour \citep{Iyyani2013}.

Summarising, the low energy component  allowed us to determine the following flow parameters, $\Gamma$,  $r_{\rm ph}$,  $r_{\rm s}$, and 
  $r_{\rm 0}$.

\subsection{Comptonised peak and the dissipation site}
\label{sec:compt}
The Comptonised spectrum is expected to extend, in the comoving frame, from the energy of the seed photons ${\cal{E}}'_i = 2.7 \, kT' \equiv  2.7 \, \theta' \, m_{\rm e}  c^2$ up to  the maximum energy that the photons can be up-scattered to ${\cal{E}}_c' = (\gamma \beta)^2 \, m_{\rm e} c^2/f$,  where $\gamma \beta$ is the  electron momentum and the factor $f $ takes into account if the electrons are in the Thompson regime or not and has a value of 1 -- $\sim 3$ \citep{Pe'er2006}; ${\cal{E}}_c'$ and ${\cal{E}}_i'$ are the temperatures of the Comptonised photons and of the seed thermal photons at the dissipation site, respectively. Here $\beta \equiv v/c$, where $v$ is the electron velocity, $c$ is the velocity of light and $\gamma$ is the Lorentz factor of the electrons.

For a relativistic Maxwell-Boltzmann distribution of the electrons, the {\it observed} spectral break at high energies, $E_{\rm c}$, therefore gives an estimate of the electron momentum,
\begin{equation}
(\gamma \beta)^2 = \frac{E_{\rm c} f }{\Gamma \, m_{\rm e}c^2} \, \tau^{2/3} 
\label{eq:1}
\end{equation} 
where the factor $\tau^{2/3} = ({r_{\rm ph}}/{r_{\rm d}})^{2/3} $ takes into account the adiabatic expansion from the dissipation site until the emission is released at the photosphere. 

\begin{figure*}
\begin{center}
\resizebox{84mm}{!}{\includegraphics{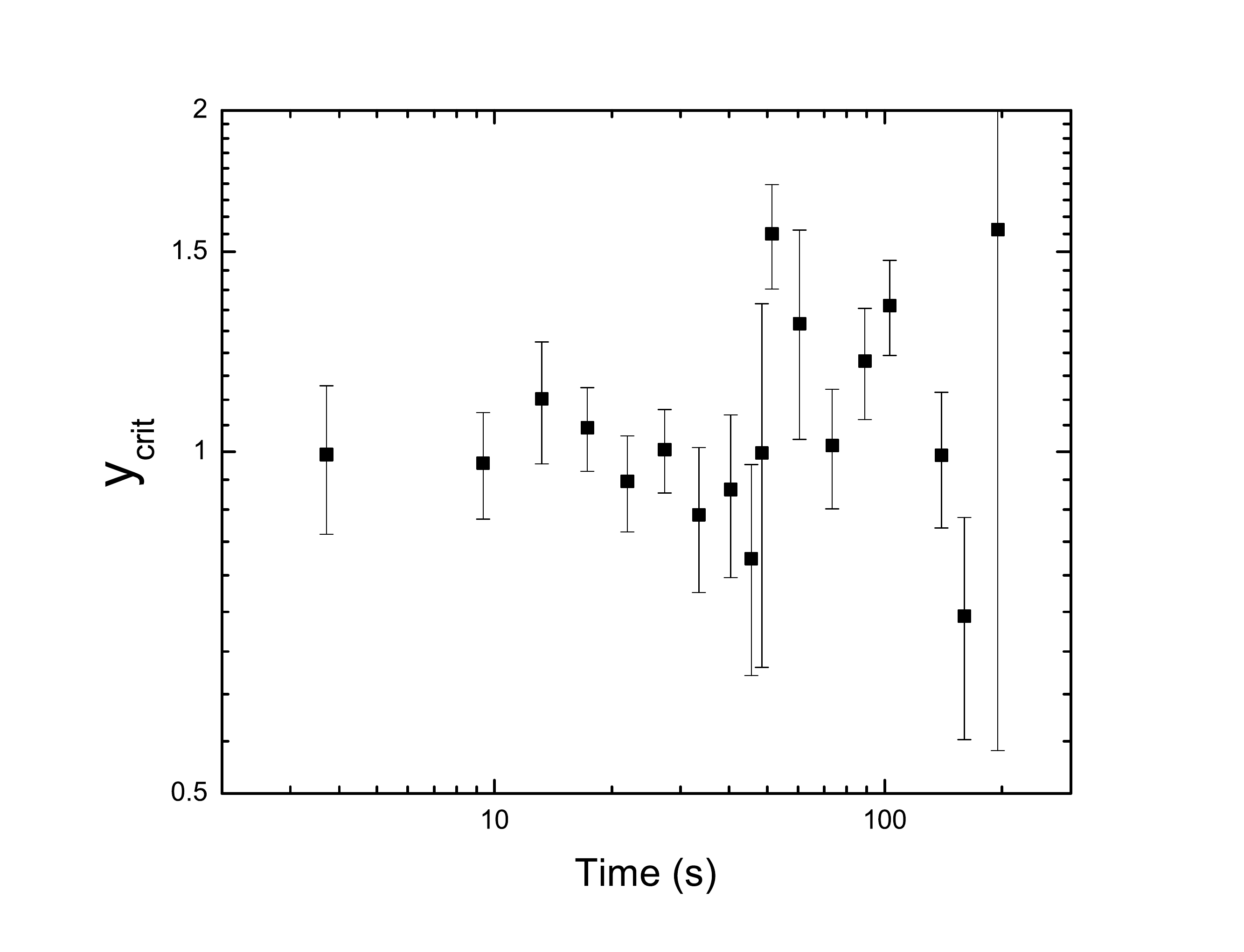}}
\resizebox{84mm}{!}{\includegraphics{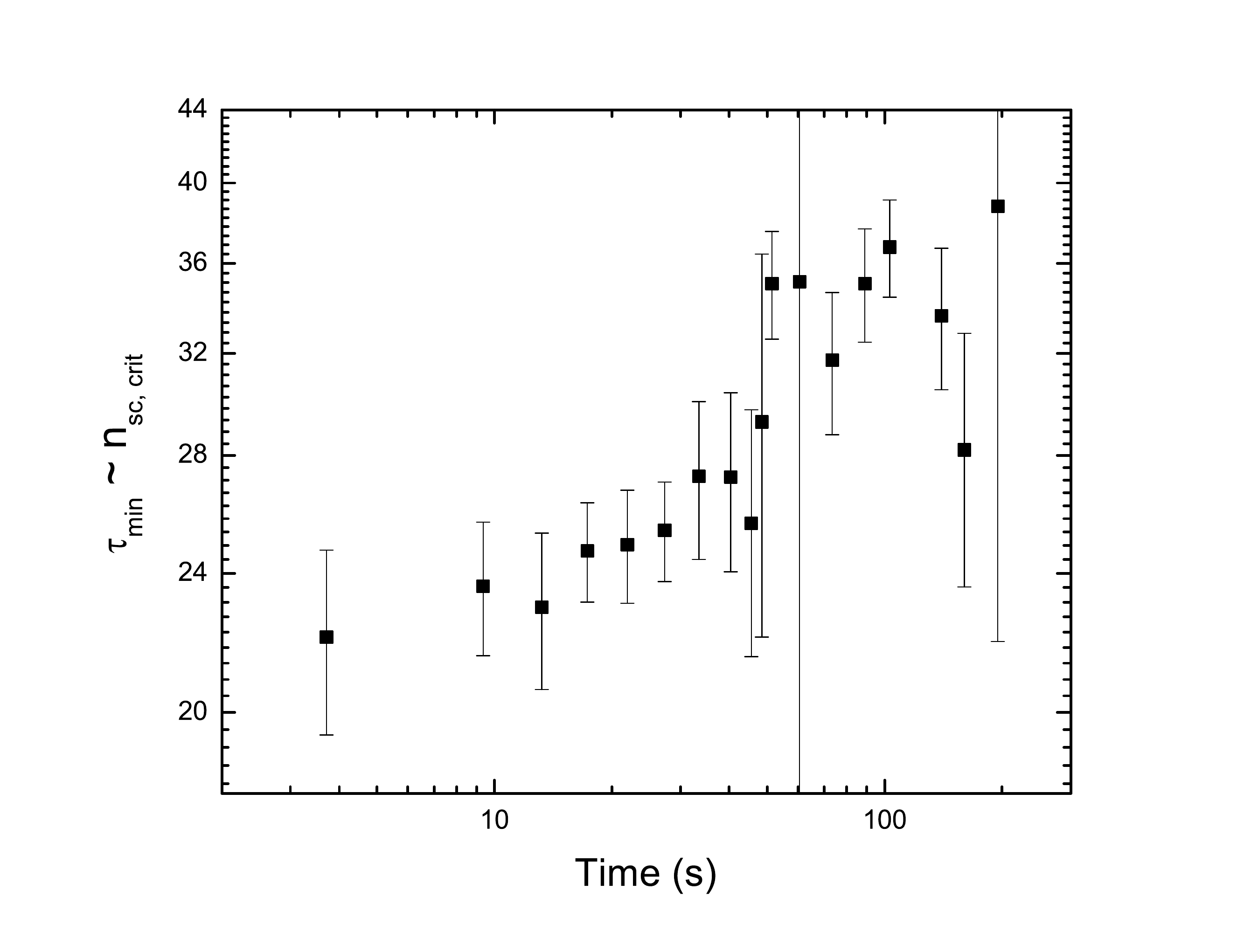}}
 
 \caption{ (a) The evolution of the critical Compton $y$ parameter, $y_{\rm crit}$. (b) The evolution of the lower limit of the optical depth at which dissipation occurs, $\tau_{\rm min} \sim n_{\rm sc, crit}$ for a $Y = 5$. }
\label{fig:inv_compt_par}
\end{center}
\end{figure*}

Furthermore, the critical Compton $y$-parameter is given by $y_{\rm crit} = \ln(E_{\rm c}/E_{\rm i})$, where $E_{\rm c}$ and $E_{\rm i}$ are the observed electron temperature and thermal, seed temperature at the photosphere. Figure \ref{fig:inv_compt_par}a shows that $y_{\rm crit}$ remains close to $1$ through out the burst. The minimum number of scatterings, $n_{\rm sc, crit}$, needed to produce the spectrum can then be estimated as
\begin{equation}
n_{\rm sc, crit} = \frac{\ln(E_{\rm c}/E_{\rm i})}{4/3 \, (\gamma \beta)^2}.
\label{eq:nsc}
\end{equation}
Since  $\tau \sim n_{\rm sc}$ for a relativistically expanding outflow, this corresponds to a lower limit to the optical depth $\tau_{\rm min} = n_{\rm sc, crit}$, which is plotted in Figure \ref{fig:inv_compt_par}b.

We can thus use equations (\ref{eq:1}) and (\ref{eq:nsc}) to estimate $\tau_{\rm min}$ and $(\gamma \beta )_{\rm min}$
by making use of the estimate of $\Gamma$ from the analysis of the seed photon component in (\S \ref{sec:cal_outflow}).
For instance, for the time bin at 10 s we find that $\tau_{\rm min}\sim 20 \, Y_{0.7}^{3/20}$ and $(\gamma \beta )_{\rm min}^2 \sim 0.03 \, Y_{0.7}^{-3/20}$ at the dissipation site\footnote{We use the notation $Y_x= Y/10^x$}. During the whole burst duration, we find that the electron momentum, $\gamma \beta_{\rm min}$ is steady around $\sim 0.17$ and thus it is apparent that the electrons, at steady state, have subrelativistic velocities (cold electrons).  
The estimated value of $\tau_{\rm min}$ corresponds to an upper limit of the dissipation radius  $r_{\rm d, max} = r_{\rm ph}/ \tau_{\rm min} \sim 8 \times 10^{10} \, Y_{0.7}^{1/10}$ cm.  The dissipation thus occurs in the vicinity of the saturation radius ($\sim 5 \times10^{10} \, Y^{-5/4}_{0.7}$ cm) such that  $   \tau_{\rm min} \sim 20 \, Y_{0.7}^{3/20} <  \tau < \tau _{\rm max} = ({r_{\rm ph}}/{r_{\rm s}}) \sim 40 \, Y_{0.7}^{3/2}$. 
The evolution of $r_{\rm d, max}$ is included in Figure \ref{fig:outflow_par}b.
Indeed,  in order for $r_{\rm d}$ to lie above $r_{\rm s}$ the above calculations indicate that  $Y \geq 4$ giving a lower limit of the value of $Y$.

The shape of the spectrum changes after $50$ s, with the spectral peak becoming narrower, and suggests an increase in  $y > y_{\rm crit}$ and correspondingly $\tau> \tau _{\rm crit}$, thereby increasing the strength of the BB2 component in the fits and creating the extreme, observed narrowness of the spectrum. The exact value of the $y$-parameter must, however, be determined by detailed modelling of the spectral shape. However, we note that since the observed spectrum does have significant deviations from a blackbody  (or a Wien spectrum), the dissipation should still be such that $r_{\rm d} \geq r_{\rm s}$. Also, $\tau$ is limited by $r_{\rm ph}/r_{\rm s} \propto L_0/\Gamma^4$. Moreover, $\tau$ must be smaller than, say, $100$ in order to avoid the creation of a pure Wien peak.   The observed spectral shape thus limits the possible values of $\tau$.

An analytical expression for the electron momentum, $(\gamma \beta)^2$ and its dependencies on the flow parameters expected for subphotospheric dissipation in GRB flows  is derived in \cite{Peer&Waxman2005}. Let us define $u_{\rm ph}$ as the initial energy density of the photon field and  $u_{\rm el}$ as the injected energy density that  the electrons have achieved due to the dissipation. Assuming that the main energy loss mechanism for the electrons is inverse Compton scattering that will balance the heating due to energy dissipation of the flow kinetic energy, a steady state value for the  electron momentum will arise 
\begin{equation}
\gamma \beta \sim \left(  \frac{9 \theta' }{4 \, \tau}    \, \frac{u_{\rm el}}{ u_{\rm ph}}    \right)^{1/4}.
\label{eq:gb}
\end{equation}

Here the ratio of ${u_{\rm el}}$ and ${ u_{\rm ph}}$ can be estimated by using the observed properties of the seed blackbody (\S \ref{sec:cal_outflow}).
If  $r_{\rm s} <  r_{\rm d} < r_{\rm ph}$ then, using the standard fireball theory \citep{Meszaros2002} and with $L_0$ being the isotropic equivalent luminosity of the burst,
\begin{equation} 
u_{\rm ph} =\frac{L_0}{4 \pi r_{\rm s}^2 \Gamma^2 c} \, \left( \frac{r_{\rm d}}{r_{\rm s}} \right)^{-8/3}
\end{equation}
and
\begin{equation} 
u_{\rm el} = \frac{L_0 \epsilon_e}{4 \pi r_{\rm d}^2 \Gamma^2 c} = u_{\rm ph} \epsilon_e  \, \left( \frac{r_{\rm d}}{r_{\rm s}} \right)^{2/3}
 \label{eq:uel}
 \end{equation}
where $\epsilon_e$ is the fraction of the dissipated energy that goes into electrons, and noting that $r_{\rm d} = r_{\rm ph}/\tau$, we have 
\begin{equation} 
\frac{u_{\rm el}}{u_{\rm ph}} = \epsilon_e \left( \frac{r_{\rm ph}}{r_{\rm s}} \right)^{2/3} \, \tau^{-2/3}  
\label{eq:EcEi}
\end{equation}

Combining equations (\ref{eq:1}), (\ref{eq:gb}), and (\ref{eq:EcEi}) we can solve for the optical depth:
\begin{equation}
\tau \sim  \left( \frac{9 \, m_{\rm e}c^2}{4 f} \right)^{{3}/{7}} \Gamma^{{3}/{7}}E_{\rm c}^{-{6}/{7}} (kT)^{{3}/{7}}\epsilon_{\rm e}^{{3}/{7}} \left( \frac{r_{\rm ph}}{r_{\rm s}}\right)^{{2}/{7}},
\label{eq:tau5}
\end{equation}
\noindent
which can be estimated from the observables, $E_{\rm c}$,  $kT$ and the derived values of $\Gamma$, $r_{\rm ph}$, and $r_{\rm s}$ (from \S \ref{sec:cal_outflow}).  This expression gives an analytical expression for the general dependencies  and gives a rough estimate of the optical depth needed for a certain steady state electron temperature. As an example the $10$ s time bin in GRB110920A gives $\tau \sim 10 \, Y_{0.7}^{15/28} $ ($\epsilon_e = 0.1$ was assumed) which is roughly consistent with the estimate from the width of the spectrum above.

Similarly,  noting that 
\begin{equation}
\theta ' = \frac{k}{m_{\rm e} c^2} \left( \frac{L_0}{4\pi r_{\rm s}^2 \Gamma^2 c a} \right)^{1/4} \left( \frac{r_{\rm d}}{r_{\rm s}}  \right)^{-2/3}
\end{equation}
equations (\ref{eq:1}), (\ref{eq:gb}), and (\ref{eq:EcEi}) can be rearranged to yield the dependencies of the ratio
\begin{equation}
\frac{E_{\rm c}}{E_{\rm i}} \propto \epsilon_{\rm e}^{1/2} \tau^{-1/2} \left( \frac{r_{\rm d}}{r_{\rm s}} \right)^{2/3} L_0^{-1/8} r_{\rm s}^{1/4} \Gamma^{1/4} 
\label{eq:final}
\end{equation}
where the dependence is strongest on $\tau$, $\epsilon_e$, and the ratio $(r_{\rm d}/r_{\rm s})$. Above we found that $r_{\rm d} \sim r_{\rm s}$ and $\tau$ to be  moderate. Therefore, equation (\ref{eq:final}) implies that dissipation parameter, $\epsilon_e$, cannot be too large  in order to explain the observed ratio of spectral peaks.   We note further that, in the case that $r_{\rm d} \sim r_{\rm s}$, equation (\ref{eq:uel}) yields that $u_{\rm el}/u_{\rm ph} \sim \epsilon_{\rm e}$.

The observed change in the shape of the spectrum after $50$ s, in which the high-energy peak becomes more prominent, requires the optical depths to increase, as discussed above. This will cause the steady state electron momentum  (eq. \ref{eq:gb}) to decrease more rapidly than the observed temperature decrease. This in turn will cause the distance between the peaks to  get smaller, as is indicated by the blue line in Figure \ref{fig:shabnam}. One possibility for the optical depth to increase is that, while $r_{\rm ph}\propto \Gamma^{-3}$ typically increases, one can imagine  $r_{\rm d}\sim 10^{11}$ cm to be relatively steady (related to the progenitor size scale). This will cause an increase in $\tau = r_{\rm ph}/r_{\rm d}$.

Summarising, the high-energy break allowed us to determine an additional parameter for the burst, namely the dissipation radius, $r_{\rm d}$.

\section{Discussion}
\label{Discussion}

Several different types of MeV spectra have been identified in time resolved analysis of  GRBs. A small fraction of the bursts are pure blackbody, others have a double peaked shape (often fitted by a Band + blackbody), while yet others are best fitted by a single Band function.

One interpretation of the different types of spectra is therefore  that, in addition to a blackbody (passive jet photosphere), a substantial part of the spectrum is formed  by processes far above the photosphere, e.g by optically-thin synchrotron emission. A varying combination of these components give rise to different spectral shapes. Another interpretation is that the spectrum is fully produced by the photosphere and the different shapes are due to varying dissipation properties below the photosphere (still allowing for dissipation above the photosphere, but not as a main ingredient). The analysis made above on GRB110920A supports the latter scenario.

 \subsection{General scenario}
 \label{gen_scenario}

The top hat spectra observed in GRB110920A  have a shape that is in between a pure blackbody spectrum, observed in {\it CGRO} BATSE \citep{Ryde2004} and {\it Fermi} bursts {\citet{Ryde2010, Ghirlanda2013,Larsson2015}  and the double-peaked spectra observed by {\it Fermi}, exemplified by GRB100724B \citep{Guiriec2011} and GRB110721A \citep{Axelsson2012}, where the ratio of the spectral break energies are typically ${E_c}/{E_i} \sim 20$  \citep{Burgess2014a}. The ratio of the spectral break energies in GRB110920A  is, however, much lower $E_{\rm c}/E_{\rm i} \sim 3$. 

Multiple spectral breaks in GRB spectra have been claimed by many studies. Using data from the PHEBUS experiment  Barat et al. (1998) found that, apart from the typical spectral break at $\sim 300$ keV, an additional break exists at around 1-- 2 MeV. Similarly, \cite{Ryde&Pe'er2009}  argues that the BATSE data (25--1800 keV) suggest additional peaks above the observed energy range (see also \cite{Battelino2007}), which was supported by observations by the EGRET instrument \citep{Gonzalez2009}. Moreover, Yu et al. (2014) A\&A, submitted, found that a spectral model including two breaks fit the data equally well as the Band function in many cases. The ratio between the breaks was found to be typically around $10$. The breaks Yu et al. (2014) introduced is for a synchrotron model, so the scenario is different from what we argue for here. However, their study indicates that the single peak identified in Band function fits can be interpreted as covering several breaks. This thus suggests that multiple breaks in GRB spectra could be more common than previously thought.

Furthermore, the Comptonised spectrum suggested above for GRB110920A, in the fits performed in this paper, is captured by an empirical model consisting of two blackbodies and a power law. Fits using a similar model have been performed on several bursts by  \cite{Basak&Rao2014, Rao2014}, however, motivated by a different scenario. In any case, these fits could be an indication that a Compotonised spectrum indeed is common and that the two spectral breaks can be identified. \cite{Basak&Rao2014, Rao2014} find the ratio between the spectral peaks to vary between $\sim3$ and $\sim 10$ in their sample. 

Based on these observed spectral properties, it can therefore be speculated that all such spectra (varying ratios of the spectral breaks) are formed primarily due to a Comptonised spectrum from a localised dissipation. Further, secondary, effects on the detailed shape of the spectra can be caused by synchrotron photons produced at low optical depths, if there are strong enough magnetic fields
\citep{Vurm2013}. 
 
The ratio between the breaks in the Comptonised spectrum is given by equation (\ref{eq:final}), with the strongest dependences on  $\tau$ and the ratio $r_{\rm d}/r_{\rm s}$. 
Since the spectral shape limits the possible variation in $\tau$, the distance between the spectral peaks are mainly set by  $r_{\rm d}/r_{\rm s}$. The closer the dissipation radius is to the saturation radius the higher the temperature of the seed photons (low-energy break)  will be since they are less affected by adiabatic losses.
Similarly, the ratio of energy density in electrons to that in the photons  $u_{\rm el}/u_{\rm ph}$ will become lower due to lower adiabatic losses. This, in turn, will cause the electron momentum (high-energy break) to decrease  (eq. \ref{eq:gb}). The distance between the two peaks will thus get smaller as the dissipation site approaches the saturation radius. 

Bursts with larger ratio between the peaks ${E_{\rm c}}/{E_{\rm i}}$, can thus be partly due to them having a different dissipation profile with a dissipation radius occurring substantially further away from the saturation radius. Indeed, for many bursts the ratio of derived  $r_{\rm ph}/r_{\rm s} $ is typically much larger than what is observed in GRB110920A and have values $r_{\rm ph}/r_{\rm s} \sim 100$ (e.g. \citet{Axelsson2012, Guiriec2011, Burgess2014a}). This yields, in turn, larger values of $r_{\rm d}/r_{\rm s}$ (assuming a certain value of $\tau$). Similarly, internal shocks are expected to occur with $r_{\rm d}/r_{\rm s} = 2 \Gamma \sim 300$.  Using such a value the ${E_{\rm c}}/{E_{\rm i}}$ ratio increases substantially (eq. \ref{eq:final}).


\subsection{Cooling of the seed photon distribution}
\label{sec:cooling}

 We pointed out above that  the temperature of the seed photons did not follow the expected power law after $\sim 50$ s (see Fig. \ref{fig:bb2pow}a). It was further argued that, since the photon number is conserved in the Comptonisation process, the seed photon distribution is smaller than it was before the dissipation episode. If the number of scattered photons is small the fitted temperature should not differ greatly from the original value, only the normalisation will decrease. However, if the number of scattered photons is large then the original distribution will be lost and the fitted value of the temperature will not correspond to the original value.
 
 In Figure \ref{fig:shabnam} the photon flux in the different components is shown. Before 50 s the number of photons in the two blackbody components are similar. However, after 50 s, the Comptonised peak dominates by a factor of four. This thus supports the argument that the erratic behaviour of the seed photon distribution, after 50 seconds (as compared to the predicted smooth power-law decay) is due to strong Comptonisation (larger $y$-parameter) leading to a loss of the possibility to measure the original temperature. It is interesting to note that for the second to last data point the fluxes are similar again and the measured temperature lines up with the power law decay, that was determined from the data points before 50 s.

\begin{figure}
\begin{center}
\resizebox{84mm}{!}{\includegraphics{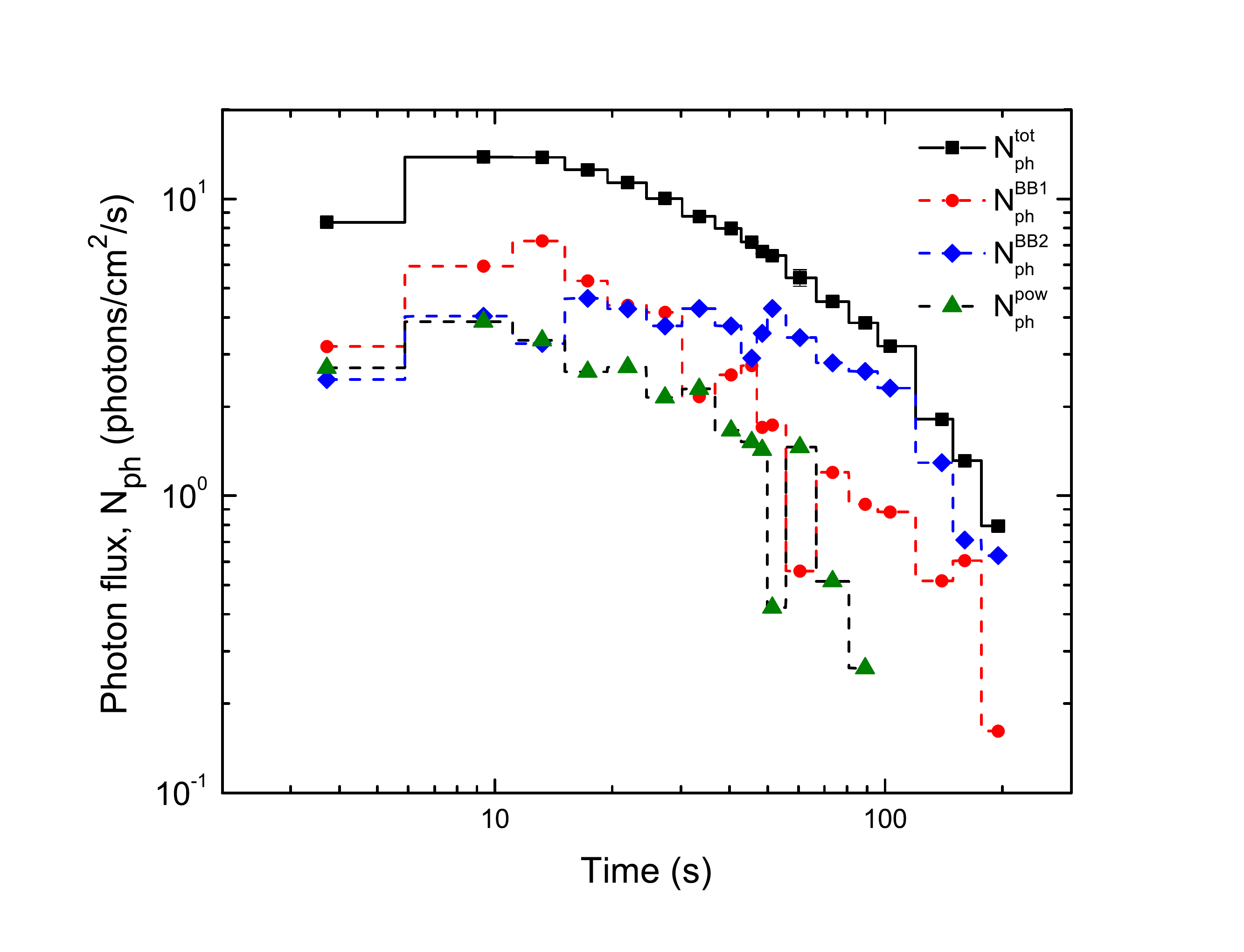}}
\caption{\small Photonflux light curves for the total flux (black/ squares), the black body components (blue/ diamond and red/ circles) and for the power law component (green/ triangles). Photons from the Comptonised peak dominated the emission after $\sim$ 50 s.}
\label{fig:shabnam}
\end{center}
\end{figure}
 
\section{Conclusion}
\label{Conclusion}

We thoroughly studied the single pulse GRB110920A since it was a strong candidate for revealing the behaviour of the photosphere.
Indeed, the extreme narrowness of the spectrum observed  directly excludes non-thermal emission models. We have demonstrated that the spectrum in GRB110920A is more complex than a Band function, having a top-hat shape with two spectral breaks.
 Such an observation restricts the amount of subphotopsheric dissipation allowed as well as the radial extent of such disturbances. If there is a large range of radii where there is strong dissipation the spectral features such as breaks will be smeared out and the resulting spectrum would be a smooth Band like spectrum. 
We find a best fit model which includes local Comptonisation which is characterised by two temperatures, the seed photon temperature and the temperature of the electrons, in addition to a weaker power-law component (which is, though, only present during the first 100 s). The main Comptonised component  enables us to determine the properties of the jet such as the bulk Lorentz factor, the photospheric radius and the dissipation radius. In particular, we find that the optical depth of the dissipation site 
is  at $\tau \sim 20$.  
The narrowness of the spectrum is mainly due to the dissipation occurring close to the saturation radius and the optical depth increasing over the burst duration. The main spectral component of GRB110920A and its evolution is thus fully explained by emission from the photosphere including localised dissipation at high optical depths.

We argue that the top hat spectra observed in GRB110920A could be the missing link between the very narrow spectra observed in many bursts (in some cases even fitted by a pure blackbody, e.g. the {\it Fermi} bursts GRB100507 and GRB101219B) and the double-peaked shape observed in {\it Fermi} bursts, such as GRB100724B and GRB110721A.Thus, in the scenario of localised subphotospheric dissipation, the main distinction between various spectral shapes is based on where the localised dissipation occurs relative to where the flow saturates. More spectral fitting using a physical model including subphotospheric dissipation is needed to assess such a framework properly.\\

\section*{Acknowledgements}

We thank the referee for useful comments on the manuscript. We acknowledge support from the Swedish National Space Board and G\"oran Gustafsson Foundation. SI is supported by the Erasmus Mundus Joint Doctorate Program by Grant Number 2011-1640 from the EACEA of the European Commission.

\bibliographystyle{mn2e}   
\bibliography{ref}

\end{document}